\documentclass[final,5p,times]{elsarticle}

\usepackage{graphics}
\usepackage{float}
\usepackage{epstopdf}
\usepackage{booktabs}
\usepackage{color}
\usepackage{subfigure}
\usepackage{graphicx}
\usepackage{eqnarray}
\usepackage{lipsum}
\usepackage{lineno}
\usepackage{amsmath,amsfonts,amsthm,bm} 

\usepackage[amssymb,thinqspace,thinspace]{SIunits}

\journal{Acta Materialia}

\begin{document}


	\begin{frontmatter}
	\title{Spinodal decomposition versus classical $\gamma'$ nucleation in a nickel-base superalloy powder: An in-situ neutron diffraction and atomic-scale analysis}

\author[label1]{David M. Collins\corref{cor1}}
\ead{d.m.collins@bham.ac.uk}
\cortext[cor1]{Corresponding author}

\author[label2]{Neil D'Souza}
\author[label3]{Chinnapat Panwisawas}
\author[label4]{Chrysanthi Papadaki}
\author[label5]{Geoff D. West}
\author[label6]{Aleksander Kostka
}
\author[label7]{Paraskevas Kontis \corref{cor2}}
\ead{p.kontis@mpie.de}
\cortext[cor2]{Corresponding author}

\address[label1]{School of Metallurgy and Materials, University of Birmingham, Edgbaston, Birmingham, B15 2TT, UK}
\address[label2]{Rolls-Royce plc, PO. Box 31, Derby DE24 8BJ, UK}
\address[label3]{NISCO UK Research Centre, School of Engineering, University of Leicester, Leicester LE1 7RH, UK}
\address[label4]{Department of Engineering Science, University of Oxford, Parks Road, Oxford, OX1 3PJ, UK}
\address[label5]{Warwick Manufacturing Group, University of Warwick, Coventry, CV4 7AL, UK}
\address[label6]{Materials Research Department and Centre for Interface Dominated Materials (ZGH), Ruhr-University Bochum, 44801, Bochum, Germany}
\address[label7]{Max-Planck-Institut f\"ur Eisenforschung GmbH, Max-Planck-Strasse 1, 40237 D\"usseldorf, Germany}

	\begin{abstract}
Contemporary powder-based polycrystalline nickel-base superalloys inherit microstructures and properties that are heavily determined by the thermo-mechanical treatments during processing. Here, the influence of a thermal exposure alone to an alloy powder is studied to elucidate the controlling formation mechanisms of the strengthening precipitates using a combination of atom probe tomography and in-situ neutron diffraction. The initial powder comprised a single-phase supersaturated $\gamma$ only; from this, the evolution of $\gamma'$ volume fraction and lattice misfit was assessed. The initial powder notably possessed elemental segregation of Cr and Co and elemental repulsion between Ni, Al and Ti with Cr; here proposed to be a precursor for subsequent $\gamma$ to $\gamma'$ phase transformations. Subsolvus heat treatments yielded a unimodal $\gamma'$ distribution, formed during heating, with evidence supporting its formation to be via spinodal decomposition. A supersolvus heat treatment led to the formation of this same $\gamma'$ population during heating, but dissolves as the temperature increases further. The $\gamma'$ then reprecipitates as a multimodal population during cooling, here forming by classical nucleation and growth. Atom probe characterisation provided intriguing precipitate characteristics, including clear differences in chemistry and microstructure, depending on whether the $\gamma'$ formed during heating or cooling.
	\end{abstract}

	\begin{keyword}
	 Superalloys \sep Neutron diffraction \sep Phase Transformation \sep Powder Metallurgy \sep Precipitation
	\end{keyword}

	\end{frontmatter}

\section{Introduction}

Understanding formation reactions and evolution behaviour of ordered second phase precipitates is technologically critical for a compendium of applications; these are most pertinent for multi-component alloys used for structural applications. This is acutely important for nickel-base superalloys, the material of choice for the hottest sections of a gas turbine engines. These alloys are designed with judicious elemental selection to confer a balance between high temperature mechanical performance, thermodynamic stability and corrosion resistance required for long-term use in an aggressive in-service environment \cite{Pollock2016,ReedSuperalloys}. The chosen multi-component compositions lead to complex microstructures and phase transformation behaviour; understanding the governing mechanisms is most critical for the selection of optimal processing parameters \cite{Pollock2006}.

Modern nickel-base superalloy turbine discs are fabricated from powders, a cost-effective processing route developed to give a step-change improvement in component performance through improved alloy homogeneity and a finer grain size over traditional cast and wrought processing methods \cite{Gessinger1974,Blackburn1977}. As there are several stages in the alloy/component manufacture, there is active research in understanding and optimising the microstructure of each processing method used. This includes hot isostatic pressing (HIP) e.g.\cite{Appa2011}, thermomechanical processing e.g.\cite{BOZZOLO20125056} and heat treatments e.g.\cite{WANG2019287}. It is notable that the majority of our understanding of microstructural control is derived from the final heat treatment final stage. The approach, however, has a clear weakness as it inhibits any understanding of $\gamma \rightarrow \gamma'$ reaction behaviour during the early stages of processing.

In polycrystalline nickel-base superalloy derivatives, the multiphase microstructure comprises an A1 structure $\gamma$ matrix and a dispersion of ordered, coherent L1$_2$ structure $\gamma'$ precipitates with a low lattice misfit. The formation of $\gamma'$ precipitates has been studied extensively in systems of varying compositional complexity, focusing on how the precipitates nucleate and grow. This has included binary Ni-Al \cite{WENDT19831649, KIRKWOOD1970563}, ternary Ni-Al-Cr \cite{Schmuck1997}, Ni-Al-Ti \cite{VOGEL2012226} and multicomponent systems  \cite{WEN20031123,PRASADRAO1987199,SINGH2013280,SINGH2011878,CHEN2015199}. Model systems have low supersaturation, making it possible to suppress $\gamma'$ formation, then observe $\gamma'$ forming via isothermal heat treatments. These studies indicate $\gamma'$ forms by classical nucleation. However, studying $\gamma'$ formation behaviour in multicomponent systems, such as those used in a commercial powder metallurgy (PM) nickel-base superalloy, contain higher concentration of $\gamma'$ forming elements, making it impossible to suppress their formation during cooling tests due to the high driving force for $\gamma'$ formation \cite{BABU20014149}. In this case, instead of classical nucleation, some authors have speculated that $\gamma'$ forms via spinodal decomposition during the early stages. Tan et al. \cite{Tan2014} showed evidence of interconnected networks of $\gamma'$ with diffuse interfaces and compositions far from equilibrium, observed following fast quenching. These results here were indicative that chemical ordering can occur concurrently to spinodal decomposition.

In order to investigate whether the formation of $\gamma'$ occurs by the classical nucleation mechanism or spinodal decomposition, we have studied the microstructural evolution of $\gamma'$ from a highly supersaturated $\gamma$ phase in a nickel-base superalloy powder. The nature of powder production sees individual particles solidify exceptionally quickly, where they are subjected to cooling rates far greater than would be possible in bulk samples, simply by virtue of the heat transfer limitations from the significant volume difference. Hence, the cooling rate during powder production is hypothesised to be sufficiently rapid to entirely suppress the $\gamma'$ formation, even in a highly alloyed superalloy derivative with high concentrations of $\gamma'$ forming elements.

In this study we have exposed a nickel-base superalloy powder to thermal cycles that reach temperatures above and below the $\gamma'$ solvus temperature whilst acquiring neutron diffraction data in-situ. Our results reveal that $\gamma'$ not only forms during cooling but also during heating via spinodal decomposition. Furthermore, depending on the heat treatment temperature, the distribution and morphology of $\gamma'$ precipitates alters from an interconnected network of precipitates, to a unimodal distribution and finally to a multimodal precipitate distribution. Understanding the precipitate formation sequence is supported by atom probe tomography measurements which ascertain characteristic compositional differences as a function of the prescribed subsolvus or supersolvus heat treatment.

\section{Method}

\subsection{Material}

Powder of the commercial nickel-base superalloy RR1000 was produced by gas atomisation with the following nominal composition (wt.\%): Ni-15Cr-18.5Co-5Mo-3Al-3.6Ti-2Ta-0.5Hf-0.06Zr-0.027C-0.015B. 

\subsection{Neutron Diffraction}

\begin{figure}[ht!]
     \centering
     \includegraphics[width=80mm]{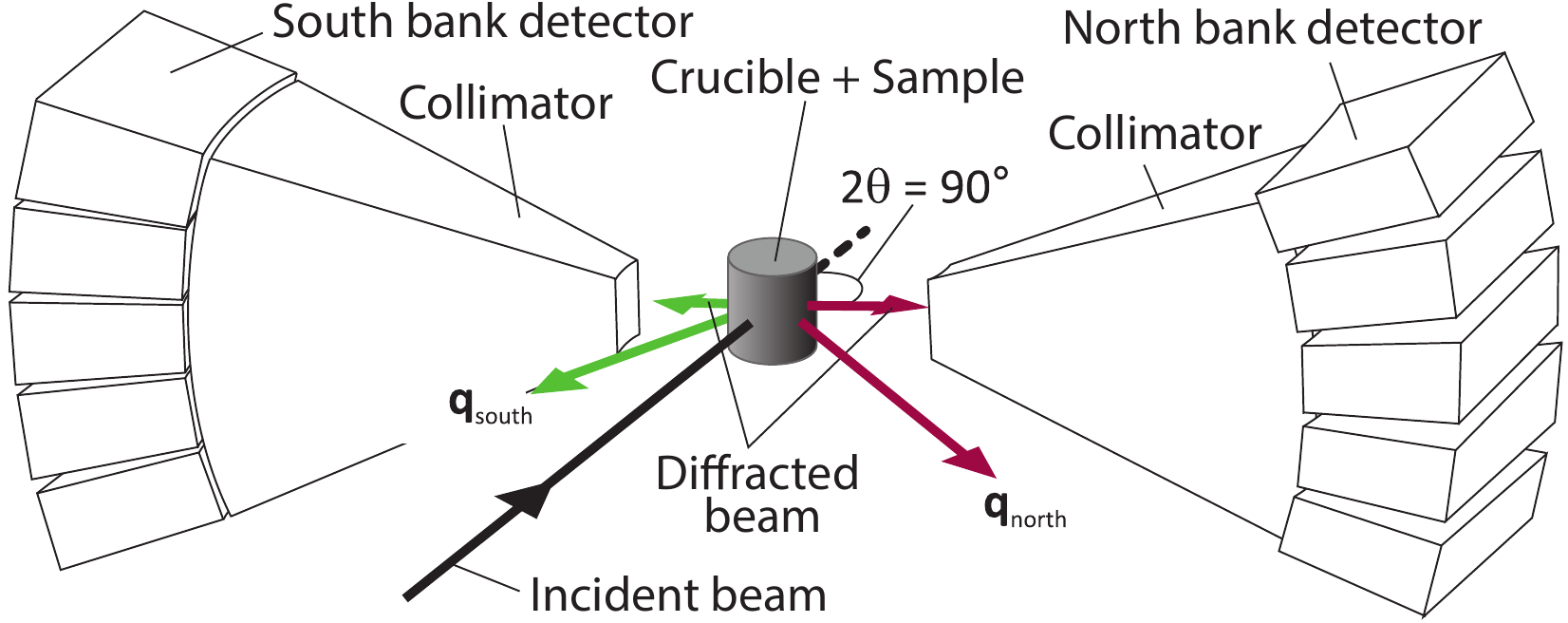}\\
     \caption{Diagrammatic experimental setup on ENGIN-X instrument, adapted from \cite{Collins_2017}.}
     \label{SetUp}
\end{figure}

Diffraction data were collected on the ENGIN-X instrument at the ISIS neutron spallation source, UK. A setup of the experiment is shown schematically in Fig. \ref{SetUp}. The beamline was equipped with a vacuum furnace suitable for in-situ heating experiments. An incident neutron beam illuminated a probed volume in the sample measuring approximately $8 \times 8 \times 8$\,mm$^3$, determined by the cross-section size of the incident mean and the collimators. The diffraction signal was measured from two detectors, both orthogonal to the incident beam, arriving from diffraction vectors parallel to the radial direction of the cylindrical sample (labelled $\bf q_{\rm{north}}$ and $\bf q_{\rm{south}}$ in Fig. \ref{SetUp}). Further instrument details are available elsewhere \cite{EnginX_2006}.

Cylindrical alumina crucibles (12\,mm outer diameter (9.5\,mm inner diameter) $\times$ 14\,mm high) were filled with alloy powder and inserted into vanadium cans before being placed into the beam line furnace. Sample temperatures were monitored using R-type thermocouples, placed in contact with the crucibles. Three thermal cycles were examined, with samples reaching (i) 900$^{\circ}$C, (ii) 1000$^{\circ}$C and (iii) 1150$^{\circ}$C. Samples were heated at  20$^{\circ}$C\,min$^{-1}$, held at the target temperature for 15 minutes, then cooled at  20$^{\circ}$C\,min$^{-1}$. The data acquisition time was 1.5 minutes per diffraction pattern.

\subsection{Diffraction Data Analysis}

From the set of diffraction patterns acquired from the time-of-flight measurements, $\gamma/\gamma'$ volume fraction, lattice parameter and lattice misfit are calculated. These parameters were obtained from the following analysis procedure, here applied to each individual diffraction pattern.

The $\gamma'$ volume faction was found by summing the integrated intensities of several $\gamma'$ superlattice reflections, comprising $\{011\}_{\gamma'}$, $\{012\}_{\gamma'}$ and $\{112\}_{\gamma'}$. Numerical integration was conducted via a simple trapezium rule for each reflection across a peak width of 0.2\,\AA. As the $(01\bar{4})_{\alpha}$ and $(116)_{\alpha}$ reflections lie in close proximity to the $\{011\}_{\gamma'}$ and $\{012\}_{\gamma'}$ reflections, a contribution of their signals was included in the numerical integration. As it was known that there was no $\gamma'$ present at $t=0$ (see APT results later) and no separate superlattice reflection contribution could be detected, the intensity associated with the alumina could be subtracted. This provides the temporal integrated $\gamma'$ intensity, $\mathbb{I}_{hkl}^{\gamma'}(t)$ for a superlattice reflection $hkl$. At the end of the experiment, at $t = t_{\rm end}$, the equilibrium $\gamma'$ volume fraction, $\phi_{eq}$, was reached. This temporal volume fraction, $\phi^{\gamma'}(t)$, could then be approximated by:

\begin{equation}
\phi^{\gamma'}(t) = \frac{\mathbb{I}_{hkl}^{\gamma'}(t)}{\mathbb{I}_{hkl}^{\gamma'}(t=t_{\rm end})}{\phi^{\gamma'}_{\rm eq}}
\label{Eq1}
\end{equation}

The $\gamma'$ lattice parameter was obtained by fitting the $\{112\}_{\gamma'}$ reflection with a Gaussian function to obtain its $d$-spacing, $d_{112}$. Assuming a powder average, the lattice parameter was given by $a^{\gamma'} = d_{112}^{\gamma'}\sqrt{N}$, where $N=6$ for the \{112\} reflection. The $\gamma$ lattice parameter was obtained by fitting a fundamental reflection associated with this phase. As the lattice parameters of the $\gamma$ and $\gamma'$ phases are near identical, their fundamental reflections overlap, necessitating separation of their independent intensity contributions. The procedure utilised here is developed from prior studies of nickel-base superalloys \cite{MA2003525, Stone19994435, COAKLEY20122729,Collins_2015, Collins_2017}. 

For a fundamental reflection, as reflections overlap, the observed intensity profile as a function of $d$-spacing, $d$, denoted as $I_{hkl}^{\gamma + \gamma'}(d)$, can be approximated by the sum of two Gaussian functions as:

\begin{equation}
\begin{split}
\label{GassDoublet}
I_{hkl}^{\gamma + \gamma'} (d) & = \frac{ I_0^{\gamma'}}{\beta^{\gamma'}\sqrt{2\pi}} \exp \left( -\frac{(d-d_0^{\gamma'})^2} {2(\beta^{\gamma'})^2} \right) \\
& + \frac{ I_0^{\gamma}}{\beta^{\gamma}\sqrt{2\pi}} \exp \left( -\frac{(d-d_0^{\gamma})^2} {2(\beta^{\gamma})^2} \right) + I_{\rm bkg}
\end{split}
\end{equation}

\noindent where  $I_0^{\gamma'}$ \&  $I_0^{\gamma}$ are amplitude scaling constants, $d_0^{\gamma'}$ \& $d_0^{\gamma}$ are the mean peak positions, corresponding to the planar $d$-spacing for each respective phase and $hkl$, $\beta^{\gamma'}$ \& $\beta^{\gamma}$ determine profile breadth and $I_{bkg}$ is the uniform background intensity.

Due to the high number of free fitting variables in Equation \ref{GassDoublet}, obtaining a numerically stable fitted solution using this formulation is near impossible. This is owing to the near identical $d$-spacings of the $\gamma$ and $\gamma'$ phases. Some of the variables can, however, be fixed, using information associated with the $\gamma'$ phase already obtained from the superlattice reflections. Firstly, $d_0^{\gamma'}$ can be inferred from $a^{\gamma'}$. In this data analysis, the $\{200\}$ reflection was used, hence $d_0^{\gamma'} = a^{\gamma'}/2$. As the ENGIN-X yields diffraction line profiles which are largely dominated by a large intrinsic peak width, it can be assumed that the contribution of line broadening from the sample is small, hence the line profiles for the $\gamma$ and $\gamma'$ phases are assumed equal and so $\beta^{\gamma'}$ = $\beta^{\gamma}$. The final assumption is that the ratio of integrated intensities of the $\gamma$ and $\gamma'$ reflections can be related to the phase volume fractions and structure factors via

\begin{equation}
\frac{\mathbb{I}_{hkl}^\gamma}{\mathbb{I}_{hkl}^{\gamma'}} \cong \frac{|F_{hkl}^\gamma|^2}{|F_{hkl}^{\gamma'}|^2} \frac{\phi^\gamma}{\phi^{\gamma'}}
\label{IntRatioEqn}
\end{equation}

\noindent where ${\mathbb{I}}_{hkl}^{\gamma}$ and ${\mathbb{I}}_{hkl}^{\gamma'}$ are the integrated intensities of the $\gamma$ and $\gamma'$ reflection, respectively, with miller indices $hkl$. $|F_{hkl}^{\gamma}|$ and $|F_{hkl}^{\gamma'}|$ are the $\gamma$ \& $\gamma'$ structure factors and  ${\phi^\gamma}$ \& $\phi^{\gamma'}$ are the $\gamma$ \& $\gamma'$ volume fractions. Equation \ref{IntRatioEqn} holds for fundamental reflections $\gamma$ \& $\gamma'$ reflections of equal $hkl$, where their $d$-spacings are similar. Using the measured ${\phi^{\gamma'}}$ from the fitted superlattice reflections, it follows $\phi^{\gamma}(t) = 1-\phi^{\gamma'}(t)$, assuming only $\gamma$ and $\gamma'$ phases are present. 

Structure factors for the $\gamma$ and $\gamma'$ phases were calculated for their A1 and L1$_2$ crystal structures. Composition predictions as a function of temperature were calculated with the {\it{Thermo-Calc}} \cite{ANDERSSON2002273} software using the {\it{Thermotech TCNI6}} thermodynamic database \cite{TCNI6}. All phases other than $\gamma$ and $\gamma'$ were suspended in these calculations. The calculation for the $\gamma$ and $\gamma'$ structure factors are described in detail elsewhere \cite{Collins_2015}, as well as site occupancies for alloy RR1000 \cite{COLLINS20137791}. These composition predictions were also used to calculate the equilibrium $\gamma'$ volume fraction, $\phi_{\rm eq}$, of 0.46 as required for Eq. \ref{Eq1}; this value is in close agreement to independent measurements of the RR1000 alloy \cite{Bagot2017}.

The use of a Gaussian line profile enables use of its analytical solution for its integrated intensity:

\begin{equation}
\begin{split}
{{\mathbb{I}}_{hkl}^{\gamma}} & = \int_{d_{\rm{min}}}^{d_{\rm{max}}} I_{hkl}^{\gamma}(d) \\
& = \frac{1}{2} I_0^{\gamma} \Bigg\{{\rm{erf}}\left(\frac{d_{\rm max}-d_0^{\gamma}}{\beta^{\gamma}}\right) {-\rm{erf}}\left(\frac{d_{\rm min}-d_0^{\gamma}}{\beta^{\gamma}}\right) \Bigg\}
\end{split}
\end{equation}

\noindent Rearranging for $I_0^{\gamma}$ and substituting Equation \ref{IntRatioEqn} gives the following relationship:

\begin{equation}
\begin{split}
\label{GaussCouple}
{{I_0}^{\gamma}}  = \left(\frac{|F_{hkl}^\gamma|^2}{|F_{hkl}^{\gamma'}|^2} \frac{\phi^\gamma}{\phi^{\gamma'}} \right) \frac{{{2 \mathbb{I}}_{hkl}^{\gamma'}}}{{\rm{erf}}\left(\frac{d_{\rm max}-d_0^{\gamma}}{\beta^{\gamma}}\right) {-\rm{erf}}\left(\frac{d_{\rm min}-d_0^{\gamma}}{\beta^{\gamma}}\right) } 
\end{split}
\end{equation}

\noindent This function was incorporated into the fitting function, Equation \ref{GassDoublet}, when fitting a $\gamma/\gamma'$ fundamental reflection. Once $d_0^{\gamma}$ was obtained and $a^{\gamma}$ was calculated, the lattice misfit, $\delta$, was obtained for each diffraction measurement using the expression:

\begin{equation}
\label{Eqn_Misfit}
\delta = 2\frac{a_{\gamma'}-a_{\gamma}}{a_{\gamma'}+a_{\gamma}}
\end{equation}

\subsection{TEM characterisation}
Powders were mounted in conductive Bakelite and polished using conventional metallographic preparation procedures with an extended final colloidal silica stage to minimise preparation induced deformation in the surface.  Samples for Transmission Electron Microscope (TEM) analysis were prepared using a FEI Versa 3D dual beam a system combining Focussed Ion Beam (FIB) and a Field Emission Gun Scanning Electron Microscope (FEG-SEM).  Samples were extracted from within a single powder particle using a standard in-situ liftout procedure, however great care was taken to ensure that the thickness of the samples was homogeneous and similar for all samples.  The composition of the phases within a powder particle were measured using a FEI Talos F200X Field Emission Gun (Scanning) Transmission Electron Microscope (FEG-(S)TEM) equipped with a Super X energy dispersive spectroscopy (EDS) system.  Chemical distribution spectra maps were collected using EDS in STEM mode and quantification was performed from reconstructing areas within the maps from multiple areas where there was no/minimal apparent overlap between the phases using Velox software.

Further TEM characterisation was conducted where diffraction contrast microscopy was used to produce electron diffraction and dark-field images. Here, a liftout procedure was used to prepare a TEM foil using an FEI Helios G4 CX FIB operated at 30\,kV. A low, 5\,kV, ion beam was used for surface cleaning for a period of 2 minutes on each side of the TEM samples in order to remove the outer layer of the specimens affected by FIB beam damage. The specimen was observed using a FEI Tecnai F20 operated at 200\,kV.

\subsection{Atom Probe Tomography}

To obtain three-dimensional compositional information of the  $\gamma$  and  $\gamma'$ microstructure, samples were prepared for atom probe tomography (APT) via a site specific lift out method (described in \cite{Thompson2007}) using an FEI Dual Beam FIB Helios 600i. Samples from the initial powder and from samples heat treated in the beam line experiments (these samples had 900$^\circ$C, 1000$^\circ$C and 1150$^\circ$C thermal exposures) were characterised. The specimens were extracted from random sites within each sample. The APT measurements were conducted using a Cameca LEAP 3000 HR instrument, operated with a pulsed laser at 200\,kHz, with a pulse energy of 0.4\,nJ and at a temperature of 60\,K. Data reconstruction and subsequent analysis was performed using the software \textit{IVAS 3.8.2}.

\section{Results}

\subsection{Starting Powder}


\begin{figure}[ht!]
     \centering
     \includegraphics[width=70mm]{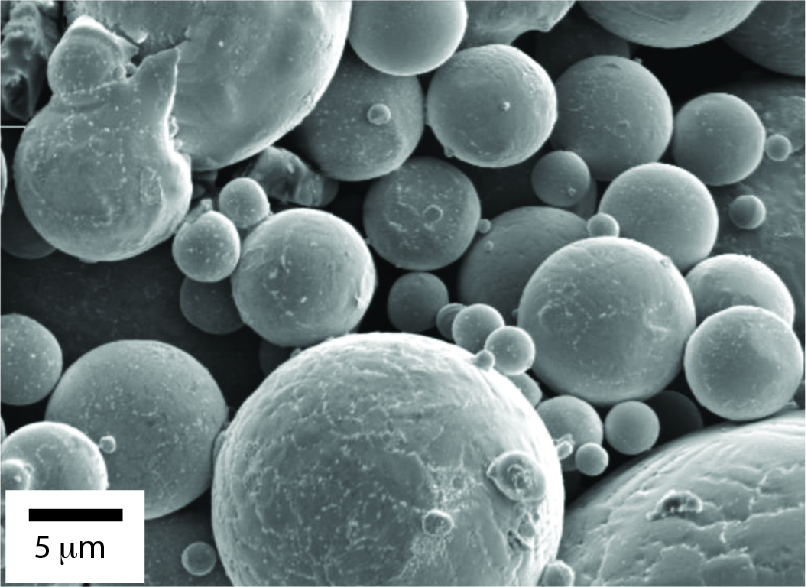}\\
     \caption{Secondary SEM micrograph of RR1000 powder after gas atomisation.}
     \label{PowderFig}
\end{figure}

A secondary electron micrograph of the starting alloy powder is shown in Fig. \ref{PowderFig} indicating the spherical morphology of the individual powder particles and a distribution of diameters, approximately in the range 2-15\micro m. Atom probe tomography measurements were made on the starting powder; composition measurements from 3 different samples, each taken from the same powder particle, are shown in Table \ref{tab:table1}. The composition was approximately uniform within each specimen without showing any significant variations. 

\begin{table}[]
  \begin{center}
    \caption{Summary of RR1000 powder composition as collected by APT (at.\%).}
    \label{tab:table1}
\begin{tabular}{@{}cccc@{}}
\toprule
\textbf{at.\%} & \textbf{Sample 1} & \textbf{Sample 2} & \textbf{Sample 3} \\ \midrule
Ni             & 50.10             & 49.65             & 50.17             \\
Al             & 6.61              & 6.62              & 6.27              \\
Ti             & 3.19              & 3.60              & 3.14              \\
Ta             & 0.25              & 0.40              & 0.40              \\
Co             & 18.66             & 18.53             & 18.9              \\
Cr             & 18.57             & 18.21             & 18.08             \\
Mo             & 2.38              & 2.78              & 2.85              \\
Hf             & 0.10              & 0.03              & 0.03              \\
B              & 0.01              & 0.02              & 0.01              \\
C              & 0.02              & 0.02              & 0.02              \\ \bottomrule
\end{tabular}
  \end{center}
\end{table}

\begin{figure}[ht!]
     \centering
     \includegraphics[width=80mm]{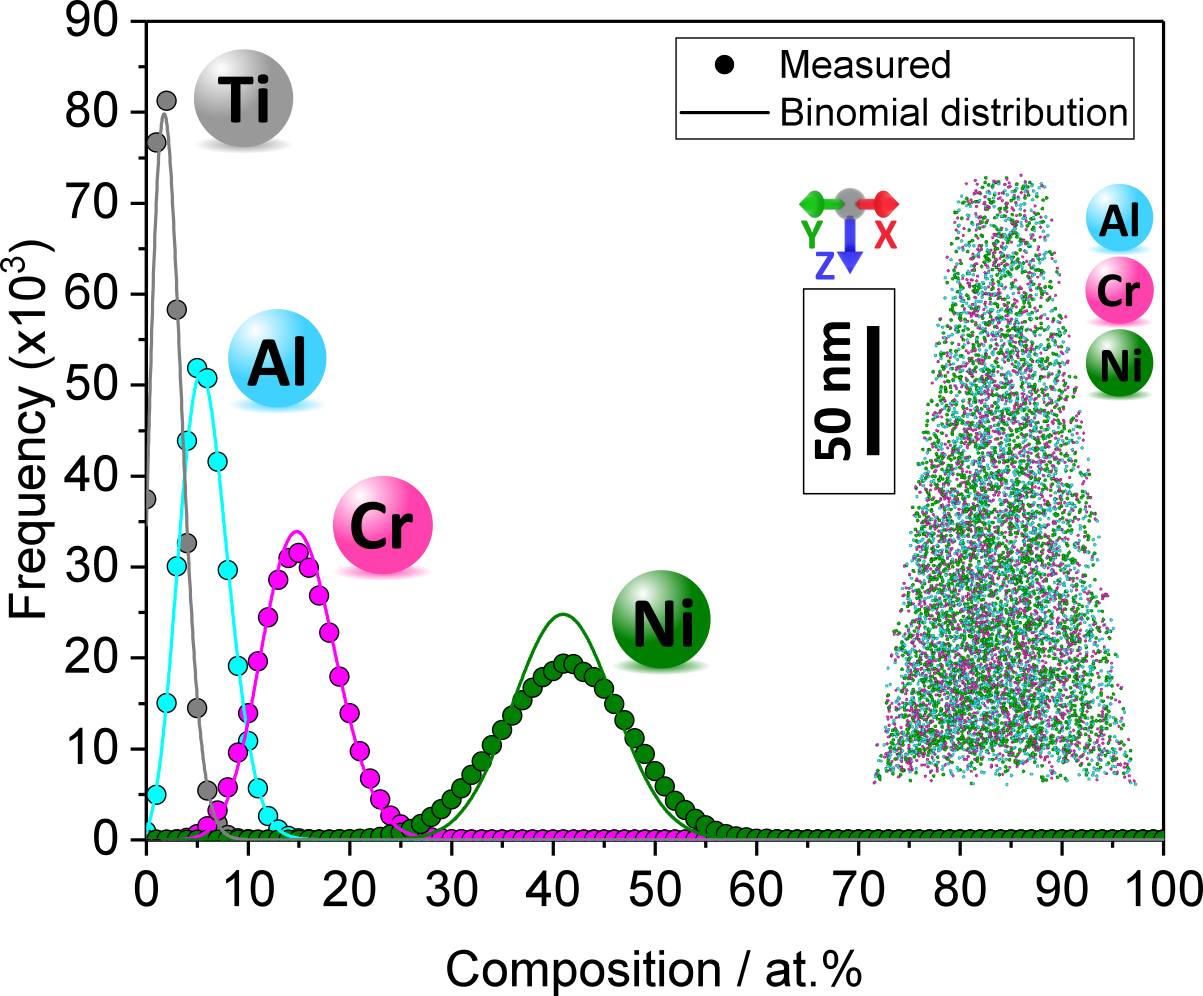}\\
     \caption{APT reconstruction from an initial powder particle alongside frequency distribution analysis of Ni, Cr, Al and Ti for the same data.}
     \label{PowderAPT}
\end{figure}

The occurrence of a compositional segregation that can be the precursor of phase separation was investigated in the initial powder. An APT reconstruction from the initial powder is shown in Fig. \ref{PowderAPT} alongside its corresponding frequency distribution analysis. In particular, the corresponding binomial, i.e. random, distribution is plotted to allow for comparison with the experimental distribution of the elements: Ni, Cr, Al and Ti. In the case of Ni, the experimental distribution does not follow the binomial distribution, indicating that Ni is not arranged completely randomly. This was further confirmed by the Pearson coefficient, $\mu$, which is given in Table \ref{tab:table2}, that can be used to reveal statistical deviation from randomness. The Pearson coefficient is indicated with values between 0 and 1, where 0 represents a random distribution and 1 reveals a statistical compositional segregation \cite{doi:10.1002/jemt.20582}. From Table \ref{tab:table2}, an obvious deviation from a random distribution is observed in the case of Ni and Cr, whereas the distribution of Ti and Al is closer to random. 


\begin{table}[]
  \begin{center}
    \caption{Pearson coefficients $\mu$ for Ni, Al, Cr and Ti as calculated for the APT data shown in Fig. \ref{PowderAPT} for the initial powder.}
    \label{tab:table2}
\begin{tabular}{@{}cccc@{}}
\toprule
  Element             & Pearson coefficient $\mu$   \\ \midrule
Ni             & 0.4539                                   \\
Cr             & 0.1331                                        \\
Ti             &  0.0478                         \\
Al              & 0.0191                                        \\ \bottomrule
\end{tabular}
  \end{center}
\end{table}

\begin{figure}[ht!]
     \centering
     \includegraphics[width=80mm]{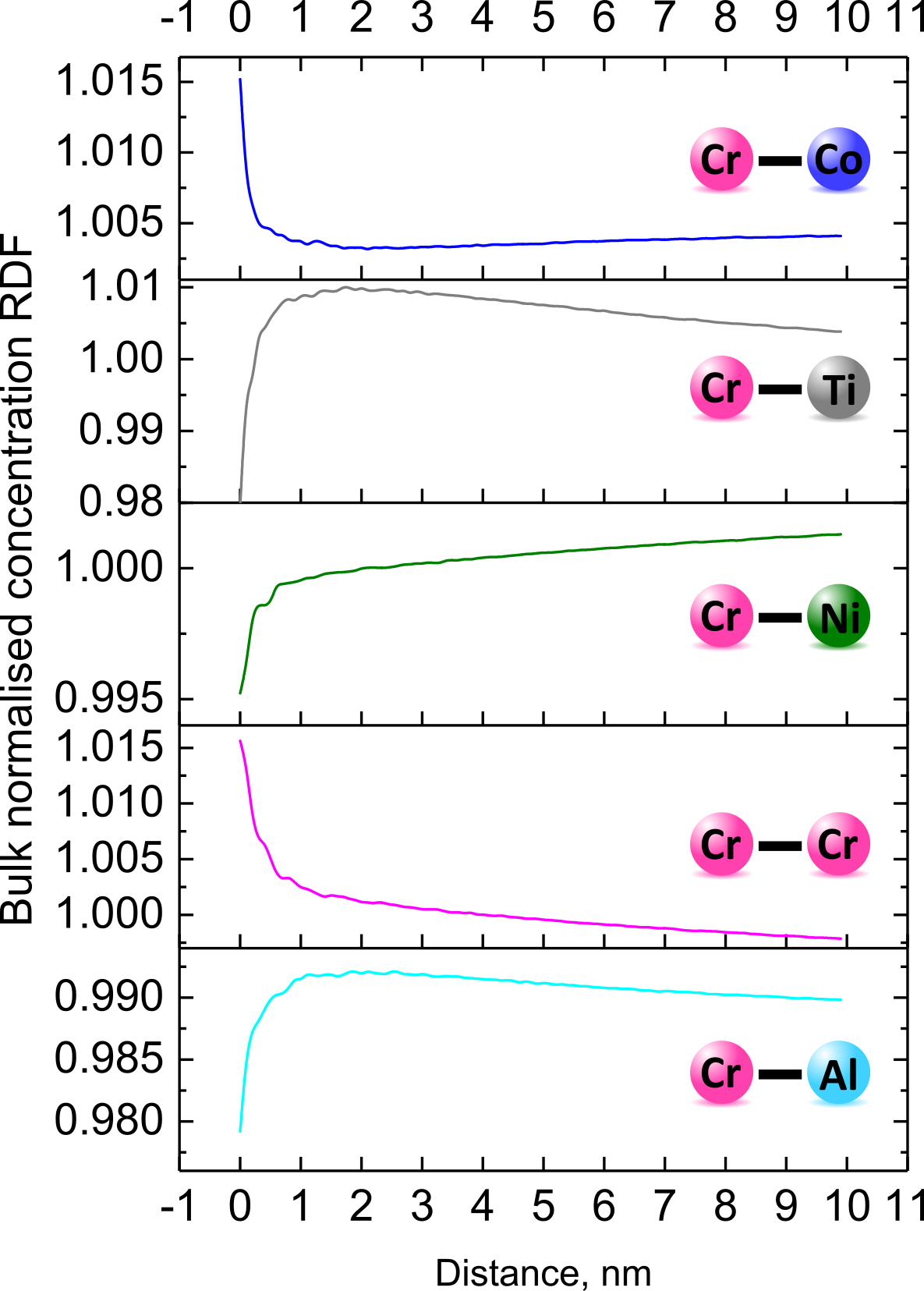}\\
     \caption{Experimental bulk normalised concentration radial distribution functions (RDFs) of Cr with Co, Ti, Ni, Cr and Al indicating a compositional segregation in the initial powder.}
     \label{RDF}
\end{figure}

In addition, radial distribution functions (RDF) were calculated from the APT data, in order to confirm a compositional segregation. In this case, each neighbourhood solute surrounding a specific species is examined and an average composition of each solute as a function of the radial distance to this particular species is calculated \cite{geuser2019metrology, PhysRevB.73.212101, Zhao_2018}). Fig. \ref{RDF} shows the RDF of Cr with Co, Ti, Ni, Cr and Al indicating a compositional segregation. In particular, the Cr-Cr and Cr-Co RDFs exhibit similar behaviour indicating that Cr and Co atoms tend to prefer to segregate together, creating regions of high Cr/Co content. In contrast, Ni, Al and Ti atoms appear repelled by the Cr atoms, resulting in regions enriched in $\gamma'$ forming elements.  In contrast, regions enriched in $\gamma'$ forming elements appear to form, as indicated by the concurrent segregation of Ni, Al and Ti atoms. Similar observations of such compositional separation were performed before in a Ni-5.2Al-14.2Cr (at.\%) alloy \cite{PhysRevB.73.212101}, whereas spinodal decomposition based on RDF analysis was shown recently in binary Fe-Cr alloys \cite{zhou2013, COUTURIER201661}.

\subsection{In-situ Neutron Diffraction}

\begin{figure*}[ht!]
     \centering
     \includegraphics[width=120mm]{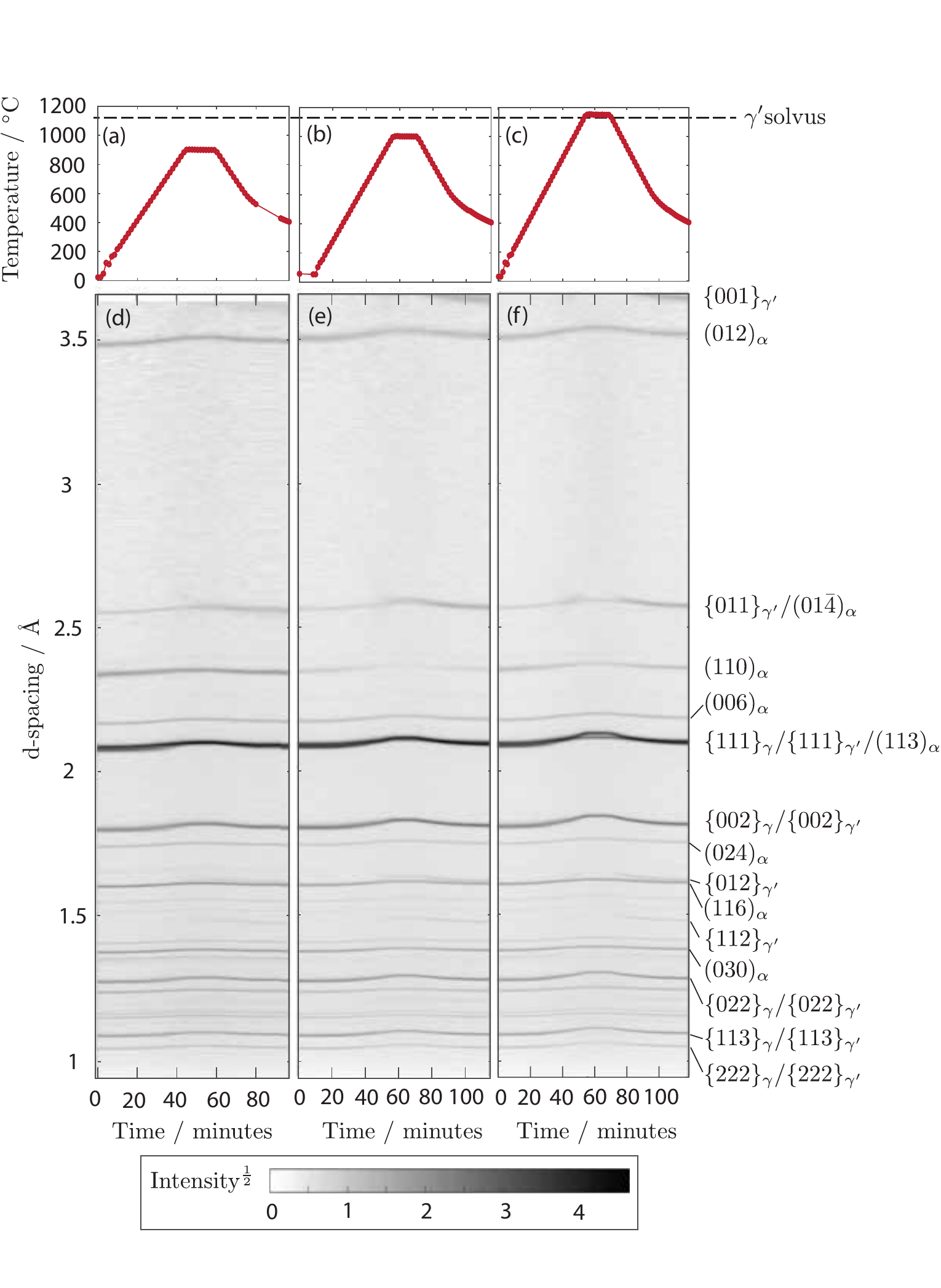}\\
     \caption{Time resolved thermocouple data and diffraction data for samples heated to 900$^\circ$C (a) \& (d),  1000$^\circ$C (b) \& (e) and 1150$^\circ$C (c) \& (f). Labelled reflections are given, the $\alpha$ planes are from the alumina crucible and $\gamma/\gamma'$ are from the matrix \& precipitation phase of the powder nickel-base superalloy.}
     \label{RawData}
\end{figure*}

The alloy powder was heated in-situ whilst collecting diffraction data. Samples heated to 900$^\circ$C, 1000$^\circ$C and 1150$^\circ$C have their respective thermocouple data shown in Fig. \ref{RawData} a, b \& c, and the corresponding temporal diffraction intensity shown in Fig. \ref{RawData} d, e \& f. The intensity data shows labels of the most intense reflections from the alloy ($\gamma$ and $\gamma'$) and the alumina crucible ($\alpha$).

\begin{figure*}[ht!]
     \centering
     \includegraphics[width=120mm]{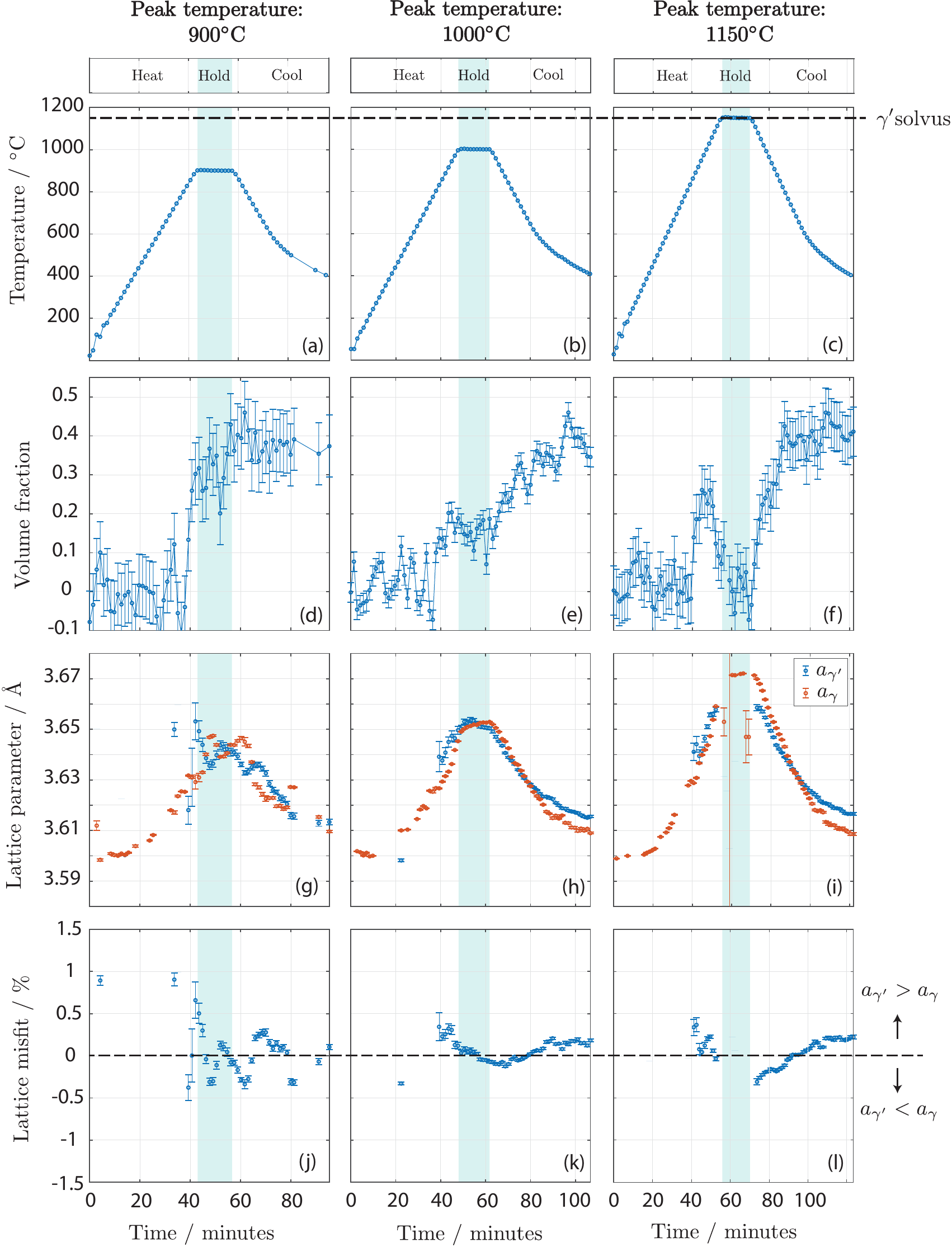}\\
     \caption{Analysed diffraction results from samples subjected to the thermal cycles shown in a, b \& c, including the corresponding volume fraction: d, e \&, f, lattice parameters of the $\gamma$ and $\gamma'$ phases: g, h \& i, and lattice misfit: j, k \& l.}
     \label{DiffractionResults}
\end{figure*}

The analysed diffraction data results are shown in Fig \ref{DiffractionResults}. Here, temporally resolved measurements of $\gamma'$ volume fraction, $\gamma$ \& $\gamma'$ lattice parameter and their corresponding lattice misfit. The key features of the differing responses for each sample will be described here in turn.

For the sample heated to 900$^\circ$C, as the powder is heated, the $\gamma'$ volume fraction remains 0\% until $\sim$700$^\circ$C. The values have been determined by tracking the evolution of $\gamma'$ superlattice reflections, giving a measure of phase ordering (and not composition localisation). Evidence of $\gamma'$ forming is clear above this temperature as its volume fraction increases; reaching $\sim$25\% once 900$^\circ$C is reached. With the short isothermal hold, the volume fraction continues to increase, reaching approximately the equilibrium volume fraction ($\sim$0.46 as calculated by {\it{Thermo-Calc}}) prior to cooling. The volume fraction does not change during cooling. The remaining $\gamma'$ following the thermal cycle has formed during the heating and isothermal hold period only. No additional $\gamma'$ formed during cooling is evident.

During heating to 1000$^\circ$C, the $\gamma'$ begins to form at $\sim$700$^\circ$C (replicating the observation when heated to 900$^\circ$C). As the temperature passes 900$^\circ$C during heating, the $\gamma'$ volume fraction drops slightly, from $\sim$25\% to $\sim$20\% and remains at this value until the end of the isothermal hold at 1000$^\circ$C. As the sample cools, the $\gamma'$ volume fraction was seen to increase, reaching an approximately equilibrium volume fraction as the temperature drops below $\sim$500$^\circ$C. It is unambiguously clear that the $\gamma'$ remaining after the thermal cycling comprised  $\gamma'$ formed both during heating and cooling.

A final sample was heated to the 1150$^\circ$C, which is above the $\gamma'$ solvus temperature. The $\gamma'$ phase forms during heating, as observed in the prior heating studies, then dissolves as the temperature approaches the solvus temperature. When the solution heat treatment temperature is reached, only the $\gamma$ phase is present. As the alloy is cooled, the intensity corresponding to the $\gamma'$ refections increases until an equilibrium volume fraction is reached by $\sim$500$^\circ$C. The $\gamma'$ phase present at room temperature in this sample evidently all formed during the cooling stage of the thermal cycle.

The $\gamma/\gamma'$ lattice parameters and corresponding lattice misfit measurements for the three heat treatment conditions is shown in Fig. \ref{DiffractionResults} (g-l). For all tests, when the volume fraction of $\gamma'$ is low ($<$10\%) there is considerable scatter in the lattice misfit and $\gamma'$ lattice parameter measurements due to uncertainty from fitting the weak $\{112\}_{\gamma'}$ superlattice reflection; these values have therefore been eliminated from the results. For precipitates formed during heating, for 900$^\circ$C and 1000$^\circ$C samples, as shown in Fig.\ref{DiffractionResults}(j) and Fig.\ref{DiffractionResults}(k), the lattice misfit was positive, whereas the formation of precipitates during cooling from 1150$^\circ$C, Fig.\ref{DiffractionResults} (l) possessed a negative lattice misfit, which approaches zero at $\sim$700$^\circ$C.


\subsection{Post mortem characterisation}

\subsubsection{Electron microscopy}

\begin{figure}[ht!]
     \centering
     \includegraphics[width=80mm]{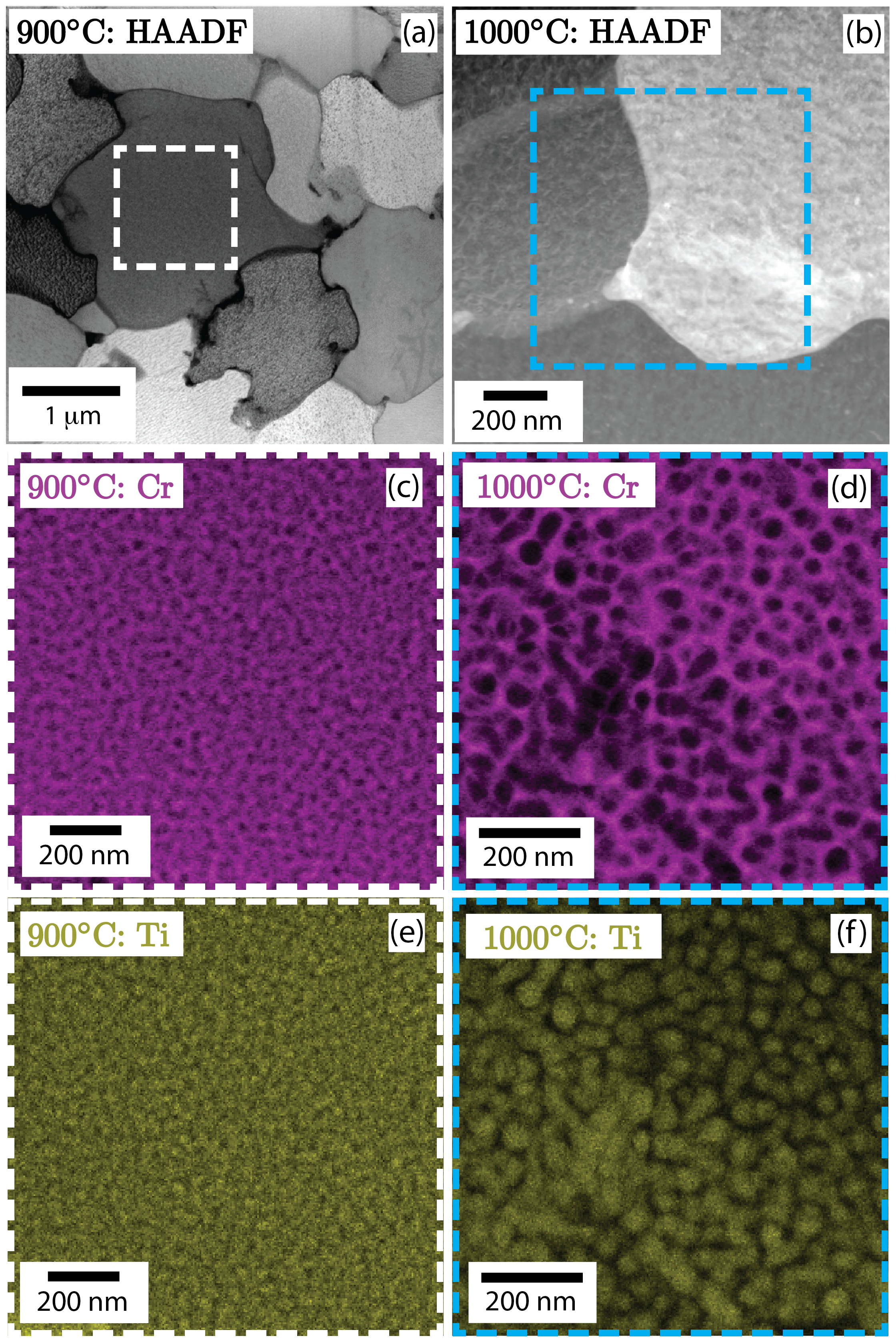}\\
     \caption{TEM micrographs of samples at 900$^\circ$C and 1000$^\circ$C (a) \& (b), with Cr maps, (c) \& (d), to indicate $\gamma$ rich regions and Ti maps, (e) \& (f), for $\gamma'$ rich regions.}
     \label{TEM_fig}
\end{figure}

TEM characterisation of the powders exposed to subsolvus $\gamma'$ temperatures were observed post mortem, as shown in Fig. \ref{TEM_fig}, with a HAADF image after 900$^\circ$C (Fig. \ref{TEM_fig} a) after 1000$^\circ$C exposure (Fig. \ref{TEM_fig} b). Neither sample was observed to have primary $\gamma'$ present at the grain boundaries. The dashed box in each HAADF image represents the composition maps for each respective sample in Fig. \ref{TEM_fig}. Cr and Ti are shown as examples of $\gamma$ and $\gamma'$ rich elements, respectively, to illustrate the size and distribution of these phases. The composition maps after 900$^\circ$C, show evidence of locally rich Ti regions, depleted of Cr, however, observing the microstructure here is resolution limited, as features appear diffuse. As a result, the morphology of individual $\gamma'$ precipitates is difficult to determine using this characterisation method. 

\begin{figure}[ht!]
     \centering
          \includegraphics[width=80mm]{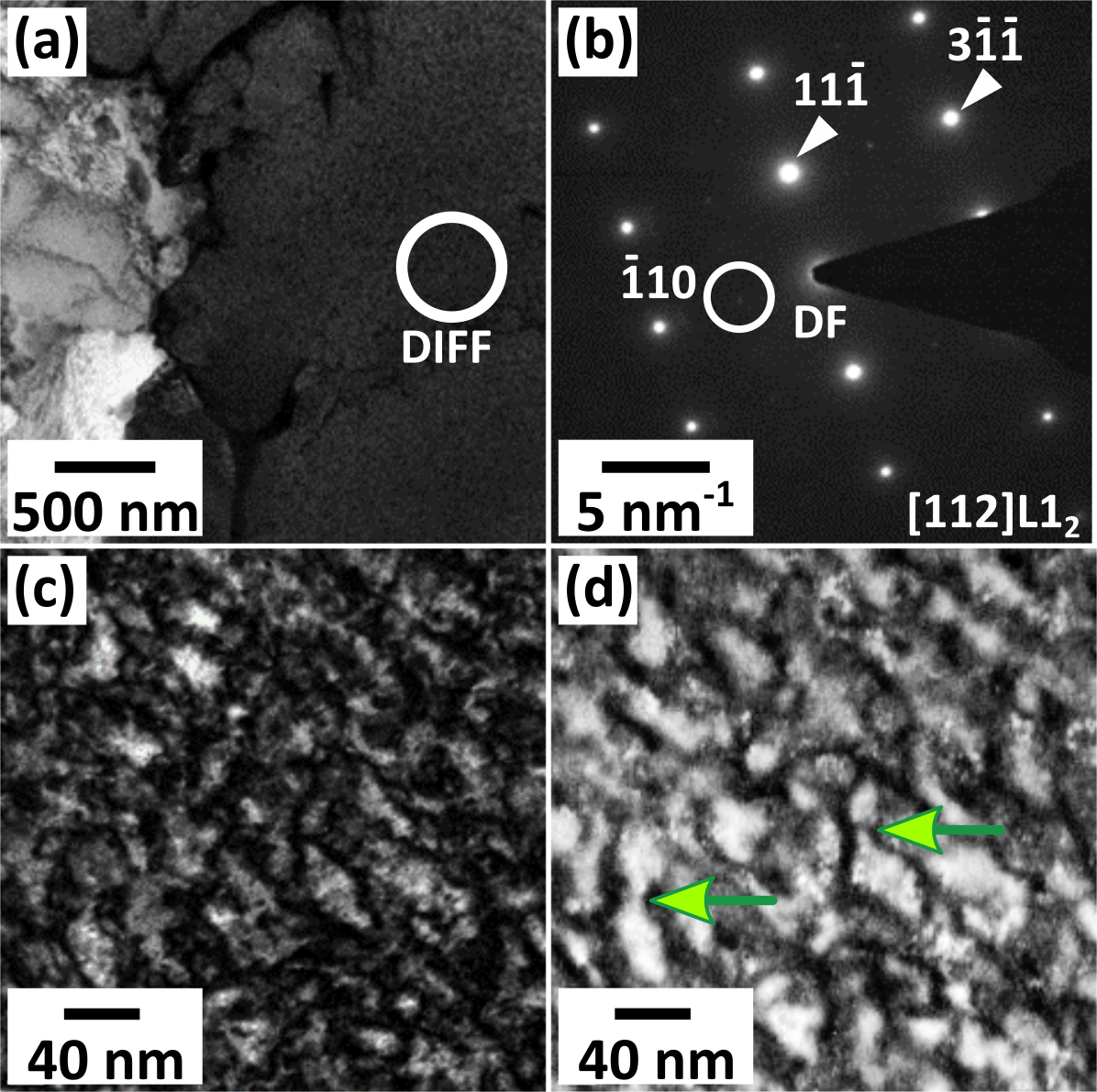}\\
     \caption{TEM micrographs showing presence of the interconnected ordered L$1_2$ phase. (a) Bright field (BF) micrograph with highlighted region where the electron diffraction pattern (b) was acquired. (b) Indexed diffraction pattern with a $[112]$ zone axis and \{${\bar{1}}1$0\} reflection selected for dark field (DF) imaging. (c, d) BF and DF micrographs corresponding to the pattern (b). The arrows in (d) denote necked regions between $\gamma'$ precipitates.}
     \label{TEM_900_diffraction}
\end{figure}

To ascertain whether the $\gamma'$ precipitates could be better determined using diffraction contrast microscopy, this sample was observed using bright and dark field imaging modes. A bright field (BF) image of the 900$^{\circ}$C annealed sample is shown in Fig. \ref{TEM_900_diffraction}a, the region highlighted as `DIFF' corresponds the region where the electron diffraction pattern (Fig. \ref{TEM_900_diffraction}b) was acquired. Using a selected superlattice (112) plane, a high magnification DF image shown in Fig. \ref{TEM_900_diffraction}d and corresponding bright field Fig. \ref{TEM_900_diffraction}c images were produced to reveal contrast between the $\gamma$ and $\gamma'$ phases. The dark/bright contrast shows clear evidence of regions of different $\gamma$ and $\gamma'$ crystallographic structures present in this sample. Referring to the DF image, a morphologically complex network of $\gamma'$ precipitates can be seen. This includes evidence of narrow necked regions (identified with arrows), appearing bright to denote an L$1_2$ $\gamma'$ crystal structure, that connect neighbouring $\gamma'$ precipitates. Some of the $\gamma'$ features are difficult to distinguish, this is because the signal is integrated through the thickness of the TEM sample of a rather complex 3D interconnected $\gamma'$ network and results in several $\gamma'$ features superimposed, that cannot be individually distinguished. Furthermore, the blurring of the $\gamma'$ is partly due to the TEM resolution limit, however, it is also possible that the $\gamma/\gamma'$ interface does not possess a sharp A1/L1$_2$ structure transition.

Inspection of the sample subjected to the 1000$^\circ$C anneal shows a population of approximately spherical $\gamma'$ precipitates can be seen (by Cr depleted regions, Fig. \ref{TEM_fig}d and Ti rich regions, Fig. \ref{TEM_fig}f), appearing as a unimodal size distribution of $\sim$40\,nm diameter.

\begin{figure}[ht!]
     \centering
     \includegraphics[width=70mm]{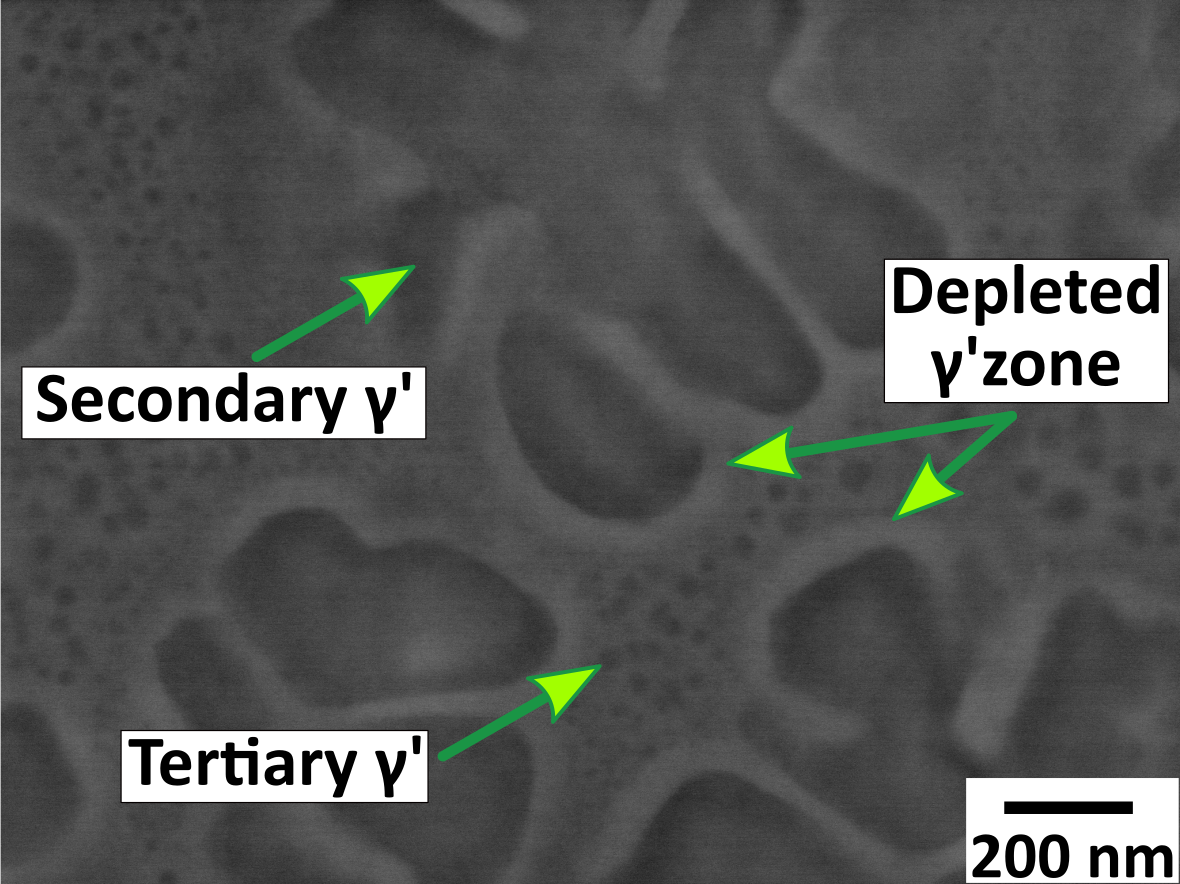}\\
     \caption{SEM micrograph after supersolvus heat treatment reaching $1150^\circ$C.}
     \label{SEM_1150}
\end{figure}

The sample subjected to a thermal cycle up to 1150$^\circ$C was examined post-mortem with a scanning electron microscope; a backscattered electron (BSE) micrograph reveals a multimodal $\gamma'$ distribution as shown in Fig. \ref{SEM_1150}. Secondary $\gamma'$ precipitates with an octodentric morphology can be seen with a size of several hundred nanometers. In between the secondary $\gamma'$ precipitates there is a much finer distribution ($< 50$ \,nm) of spherical tertiary $\gamma'$ precipitates. There is also evidence of a depleted $\gamma'$ zone in the $\gamma$ matrix in regions close to the secondary $\gamma'$ interface. These observations are near identical to the microstructure observed in fully processed RR1000 subjected to a supersolvus heat treatment followed by a slow cool (1$^\circ$C\,min$^{-1}$) \cite{Connor2014}.

From HAADF-STEM, STEM-EDS and APT characterisation results presented in the associated manuscript, this sample comprised $\gamma$ and $\gamma'$ phases with interconnected networks of $\gamma'$ precipitates, which was inferred by inspection/reconstruction of the chemical composition spacial distributions. By conducting diffraction contrast spectroscopy, the observed microstructure can be related to the crystallography of these phases.

\subsubsection{Atom probe tomography}
\begin{figure}[ht!]
     \centering
     \includegraphics[width=80mm]{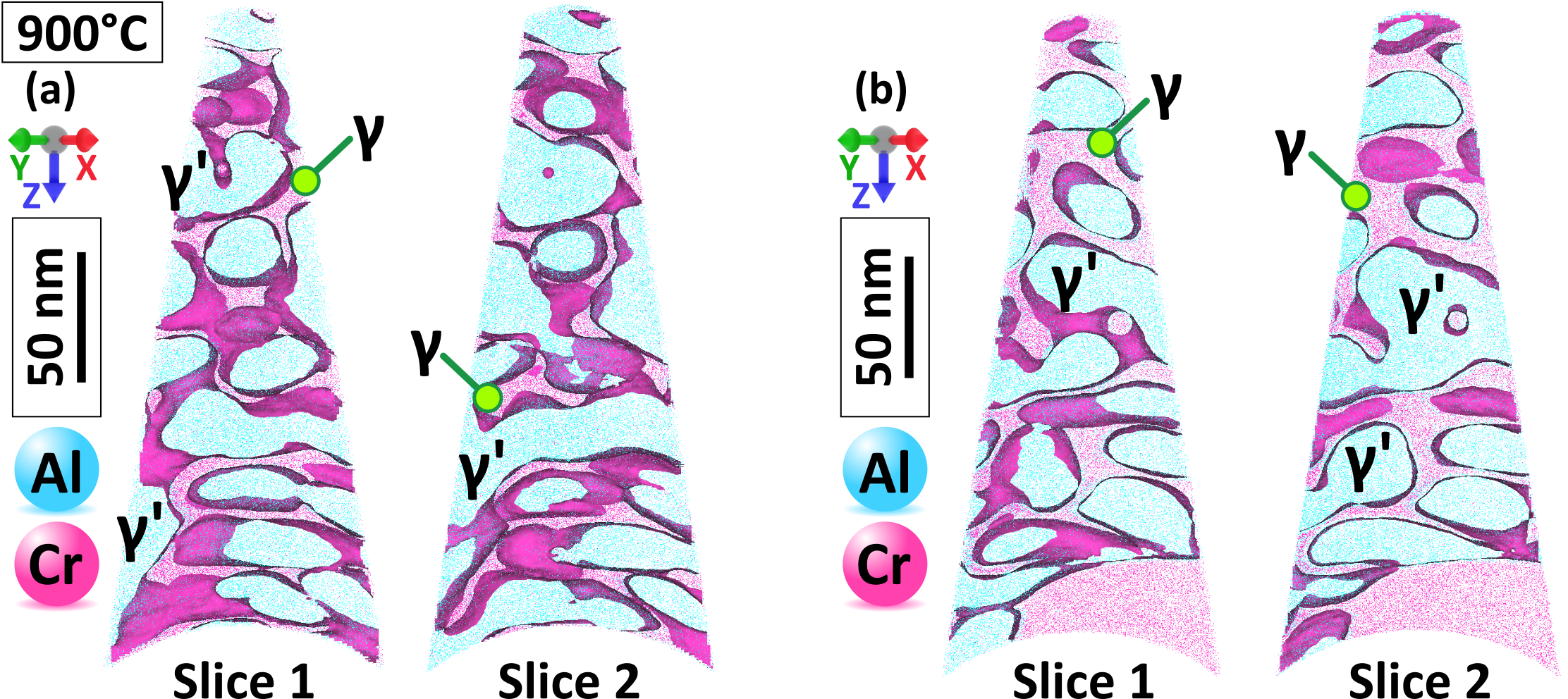}\\
     \caption{a) and b) APT reconstructions from the RR1000 powder exposed at 900$^\circ$C for 15 minutes, showing interconnected $\gamma'$ precipitates with an isosurface at 7.30 and 9.30 at.\% Cr, respectively. }
     \label{APT900}
\end{figure}

The composition of the $\gamma'$ precipitates formed at 900$^\circ$C, 1000$^\circ$C and 1150$^\circ$C was measured by APT. Fig. \ref{APT900} shows sections from two APT reconstructions from the powder exposed at 900$^\circ$C for 15 minutes. For both datasets, clear interfaces between $\gamma'$ precipitates and $\gamma$ matrix can be seen. However, the $\gamma'$ precipitates exhibit a rather complex morphology. In particular, the $\gamma'$ precipitates appear to be interconnected and have irregular shape, corroborating the observations made with TEM (Fig. \ref{TEM_900_diffraction}). Such morphology was observed previously for $\gamma'$ precipitates formed by spinodal decomposition during cooling \cite{Tan2014}. We have presented only sections of the 3D-reconstructions in order to reveal the complex microstructure. The composition of the $\gamma'$ precipitates and the $\gamma$ matrix regions as measured by APT are shown in Table \ref{tab:table900}. Although, the composition of the $\gamma'$ particles formed during heating is relatively constant, the composition of the $\gamma$ matrix shows some variations particularly in the case of Al, Ti and Co. 

\begin{table*}[h]
  \begin{center}
    \caption{Summary of $\gamma$ and $\gamma'$ composition of RR1000 powder after heating at 900$^\circ$C for 15 minutes as collected by APT (at.\%).}
    \label{tab:table900}
\begin{tabular}{@{}ccccccccc@{}}
\toprule
\textbf{at.\%} & \textbf{$\gamma$} & \textbf{$\gamma$} & \textbf{$\gamma$} & \textbf{$\gamma$} & \textbf{$\gamma'$} & \textbf{$\gamma'$} & \textbf{$\gamma'$} & \textbf{$\gamma'$} \\ \midrule
Ni             & 34.23             & 32.25             & 33.50   &  32.65 & 62.58 & 63.17 & 63.06 & 62.54       \\
Al             & 1.49            & 0.63              & 1.19      & 0.91  & 11.26 & 11.16 & 12.78 & 12.62         \\
Ti             & 1.13              & 0.56              & 0.81    & 0.54  & 13.60 & 13.06 & 12.09 & 13.06           \\
Ta             & 0.00              & 0.09              & 0.17     & 0.00  & 1.45 & 1.42 & 1.41 & 1.34          \\
Co             & 26.94             & 29.12             & 27.82    & 27.83  & 8.34 & 8.48 & 8.06 & 7.91           \\
Cr             & 33.00             & 33.83             & 32.62    & 34.39  & 1.47 & 1.47 & 1.45 & 1.36          \\
Mo             & 3.16              & 3.52              & 3.65     & 3.64  & 0.82 & 0.74 & 0.92 & 0.90          \\
Hf             & 0.00              & 0.00              & 0.02     & 0.00  & 0.41 & 0.37 & 0.14 & 0.26          \\ \bottomrule
\end{tabular}
  \end{center}
\end{table*}

\begin{figure}[ht!]
     \centering
     \includegraphics[width=90mm]{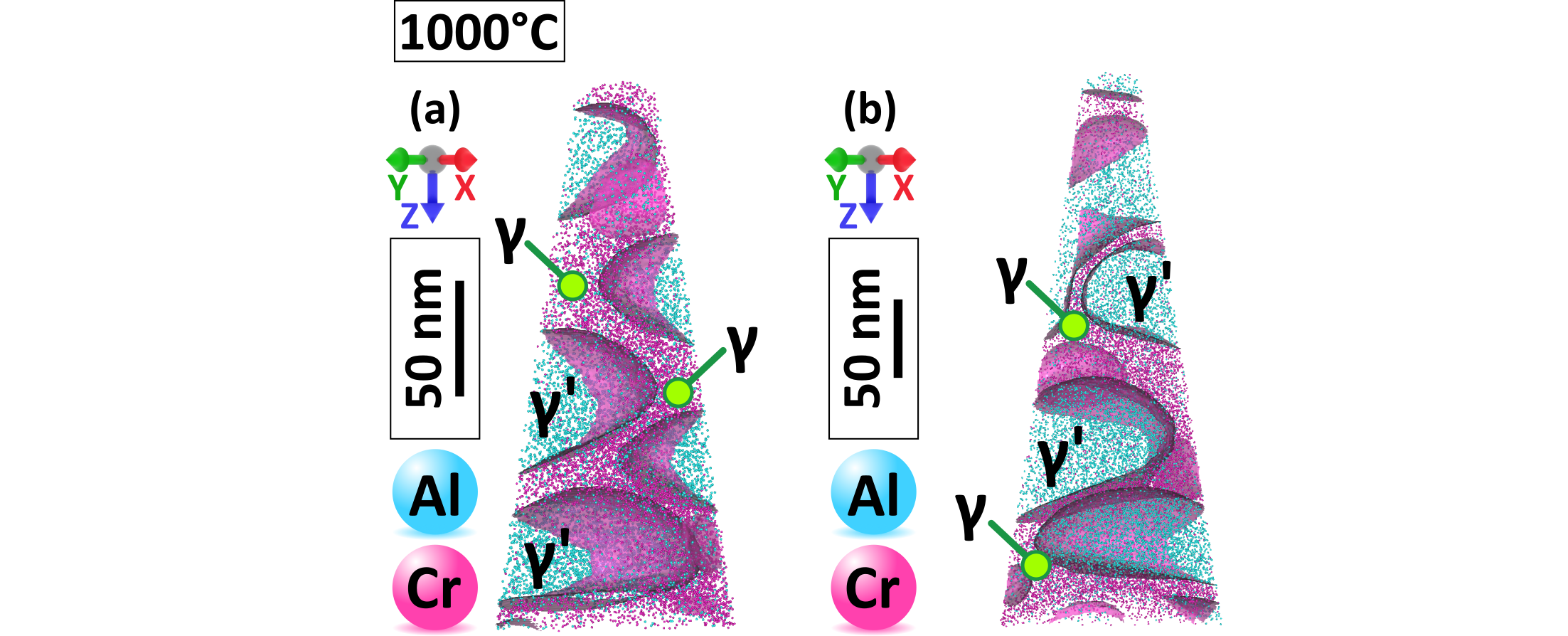}\\
     \caption{a) and b) APT reconstructions from the RR1000 powder exposed at 1000$^\circ$C for 15 minutes, showing secondary $\gamma'$ precipitates with an isosurface at 8.00 and 9.00 at.\% Cr, respectively.}
     \label{APT1000}
\end{figure}

Fig. \ref{APT1000} shows two APT reconstructions from the powder exposed at 1000$^\circ$C for 15 minutes. The morphology of the $\gamma'$ precipitates is different compared to that after 15 minutes at 900$^\circ$C. In particular, the $\gamma'$ precipitates are not interconnected and this is consistent with the observations from TEM in Fig. \ref{TEM_fig}. In Table \ref{tab:table1000} , the composition of the $\gamma'$ precipitates and the $\gamma$ matrix is given as measured by APT. No particular variation across the compositions is observed. However, the composition of the of the $\gamma'$ precipitates and the $\gamma$ matrix measured at 1000$^\circ$C shows some variation when compared to that of 900$^\circ$C.

\begin{table*}[h]
  \begin{center}
    \caption{Summary of $\gamma$ and $\gamma'$ composition of RR1000 powder after heating at 1000$^\circ$C for 15 minutes as collected by APT (at.\%).}
    \label{tab:table1000}
\begin{tabular}{@{}ccccccccc@{}}
\toprule
\textbf{at.\%} & \textbf{$\gamma$} & \textbf{$\gamma$} & \textbf{$\gamma$} & \textbf{$\gamma$} & \textbf{$\gamma'$} & \textbf{$\gamma'$} & \textbf{$\gamma'$} & \textbf{$\gamma'$} \\ \midrule
Ni             & 39.32             & 38.95            & 39.25   &  38.55 & 66.83 & 65.62 & 65.49 & 65.77      \\
Al             & 2.10           & 2.01              & 2.14      & 1.87  & 11.27 & 12.61 & 12.95 & 12.31         \\
Ti             & 0.64             & 0.66              & 0.58    & 0.49  & 9.29 & 9.90 & 10.35 & 10.08           \\
Ta             & 0.05              & 0.06              & 0.01     & 0.07  & 1.31 & 1.33 & 1.02 & 1.32          \\
Co             & 24.60            & 24.61             & 25.78    & 25.96  & 8.29 & 7.69 & 7.46 & 7.68           \\
Cr             & 28.86             & 29.15             & 28.3    & 28.67  & 1.74 & 1.73 & 1.73 & 1.69          \\
Mo             & 4.29              & 4.41              & 3.71     & 4.19  & 0.87 & 0.78 & 0.78 & 0.81          \\
Hf             & 0.00              & 0.00              & 0.00     & 0.00  & 0.32 & 0.27 & 0.17 & 0.28          \\ \bottomrule
\end{tabular}
  \end{center}
\end{table*}

\begin{figure}[ht!]
     \centering
     \includegraphics[width=80mm]{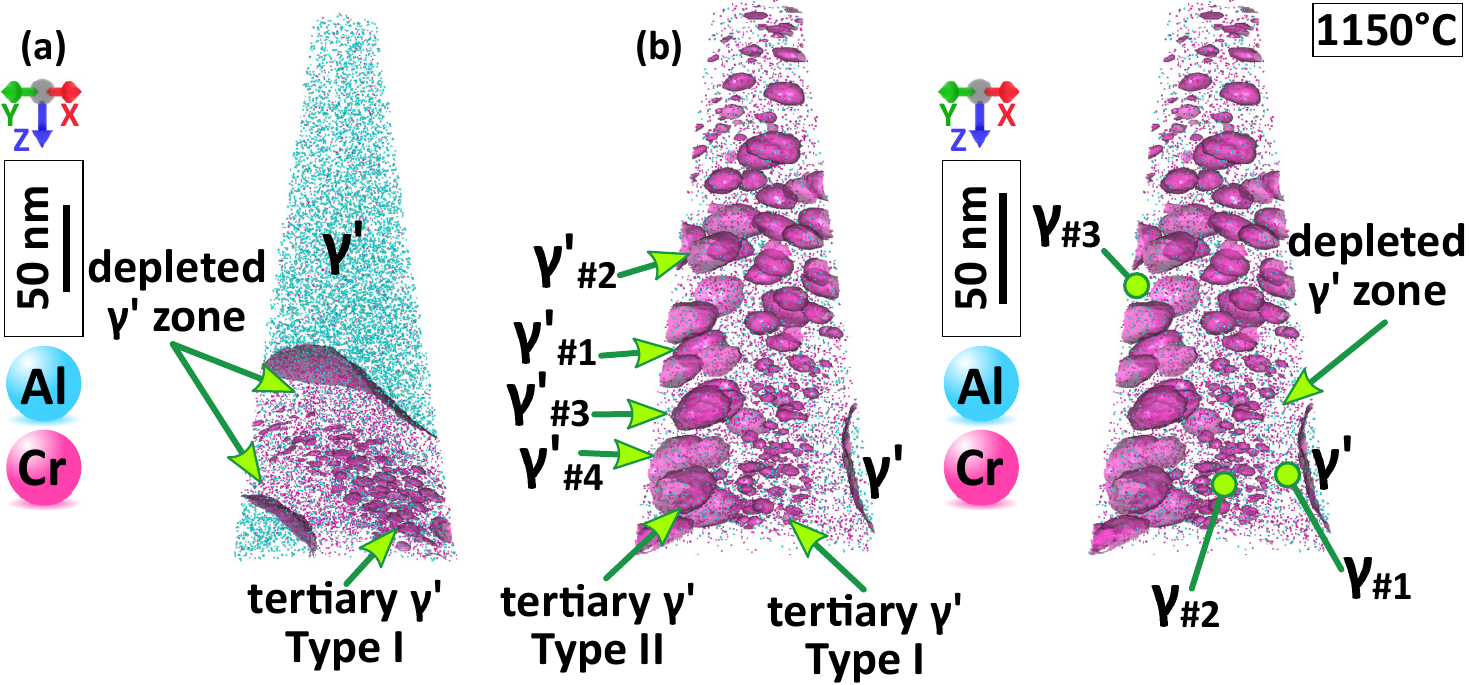}\\
     \caption{a) and b) APT reconstructions from the RR1000 powder exposed at 1150$^\circ$C for 15 minutes, showing secondary and tertiary $\gamma'$ precipitates with an isosurface at 18.00 and 19.50 at.\% Cr, respectively.}
     \label{APT1150}
\end{figure}

Fig. \ref{APT1150} shows APT 3D-reconstructions from the powder exposed at 1150$^\circ$C for 15 minutes. In particular, in Fig. \ref{APT1150}a the reconstruction contains secondary $\gamma'$ precipitates, $\gamma$ matrix and tertiary $\gamma'$ precipitates. The tertiary $\gamma'$ precipitates will be denoted as $\gamma'_{I}$ (type I). Regions depleted of tertiary $\gamma'$ precipitates in the vicinity of the secondary $\gamma'$ precipitates can also be observed. In Fig. \ref{APT1150}b, the APT reconstruction contains a part of a secondary $\gamma'$ precipitate and tertiary $\gamma'$ precipitates. In this case, two different populations of tertiary $\gamma'$ precipitates can be observed based on their size. In particular, $\gamma'_{I}$ tertiary precipitates next to a depleted $\gamma'$ precipitates zone, similar to those seen in Fig. \ref{APT1150}a. At larger distances from the depleted region, the tertiary $\gamma'$ precipitates with relatively larger size than those of $\gamma'_{I}$ can be observed. Those tertiary precipitates are denoted as $\gamma'_{II}$ (type II) tertiary precipitates. Precise quantitative information on the size of the tertiary precipitates cannot be extracted by the APT reconstructions, because the reconstructions are not based on crystallographic information that would allow that. Thus, the information on the size of the tertiary precipitates is only qualitative.

\begin{figure}[ht!]
     \centering
          \includegraphics[width=70mm]{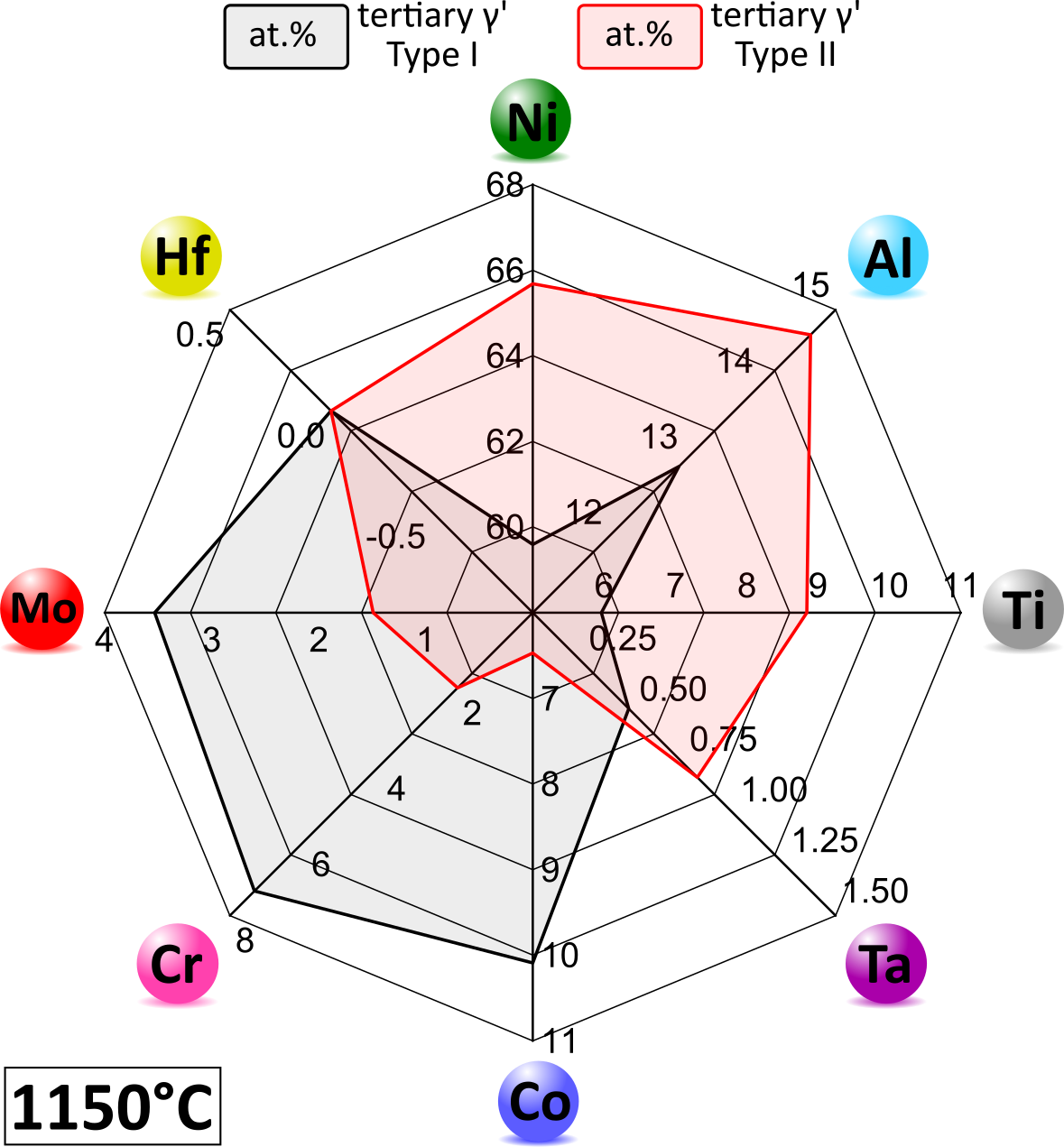}\\
     \caption{Radar-plot showing the compositional difference of between the two populations of tertiary $\gamma'$ precipitates formed after exposure at 1150$^\circ$C for 15 minutes.}
     \label{1150tertiary}
\end{figure}

The two different populations of tertiary precipitates have also significant variations in terms of chemistry as it is shown in Fig. \ref{1150tertiary}. The composition of the tertiary $\gamma'_{II}$ precipitates in this graph is the average of the four precipitates shown in Fig. \ref{APT1150}b, whereas the composition of the tertiary $\gamma'_{I}$ is the average of ten precipitates. Substantial differences can be seen for all elements between the two populations. The tertiary $\gamma'_{I}$ have higher amounts of $\gamma$ former elements, such as Cr, Co and Mo indicating that they were the last precipitates to form during cooling. As a consequence, there was not enough time to reach the equilibrium by rejecting the $\gamma$ former elements. In addition, a chemical fluctuations were observed in the $\gamma$ matrix. In particular, the $\gamma$ composition was extracted from three different locations denoted as \#1, \#2 and \#3 in Fig. \ref{APT1150}b and they are given in Table \ref{tab:table1150gamma}. It can be seen, that close to the interface of matrix with the secondary $\gamma'$ precipitate (area \#1) the amount of Cr and Co are relatively higher compared to the regions where tertiary precipitates have formed. This region correspond to the denuded region that observed close to the secondary $\gamma'$ precipitates in the powder exposed at 1150$^\circ$C for 15 minutes. By contrast, the amount of Ni is lower in the depleted region compared to the regions where tertiary $\gamma'$ precipitates have formed. It is believed that the local chemical fluctuations are due to coarsening of the secondary $\gamma'$ precipitates, that they reject $\gamma$ former elements and consume $\gamma'$ former solutes from the surrounding matrix.


\begin{table}[h]
  \begin{center}
    \caption{Summary of $\gamma$ composition of RR1000 powder after heating at 1150$^\circ$C for 15 minutes as collected from three different areas of the APT data in Fig. \ref{APT1150} (at.\%).}
    \label{tab:table1150gamma}
\begin{tabular}{@{}cccc@{}}
\toprule
\textbf{at.\%} & \textbf{$\gamma_{\# 1}$} & \textbf{$\gamma_{\# 2}$} & \textbf{$\gamma_{\# 3}$}   \\ \midrule
Ni             & 37.45             & 40.79            & 41.14         \\
Al             & 1.71           & 2.17              & 2.56               \\
Ti             & 0.5             & 0.0.75              & 0.68               \\
Ta             & 0.00              & 0.08              & 0.10               \\
Co             & 26.68            & 22.98             & 24.19               \\
Cr             & 29.14             & 27.89             & 26.85              \\
Mo             & 4.39              & 5.19              & 4.40               \\ \bottomrule
\end{tabular}
  \end{center}
\end{table}

\section{Discussion}

This study newly reveals the $\gamma \rightarrow \gamma'$ transformation behaviour from a Ni-base superalloy powder, building on knowledge obtained from numerous studies that targets the understanding of the equivalent transformation behaviour, but instead using material that has undergone no prior thermo-mechanical processing. By removing the thermo-mechanical processing aspects, this study aimed to identify how the $\gamma \rightarrow \gamma'$ evolution changes, when such influences are removed. In the following discussion, the mechanisms associated with precipitate formation during heating and cooling are discussed, followed by its relevance to practical application and an assessment of the characterisation methods used in this study.
 
\subsection{Precipitate formation during heating}

\begin{figure*}[ht!]
	\centering
	\includegraphics[width=120mm]{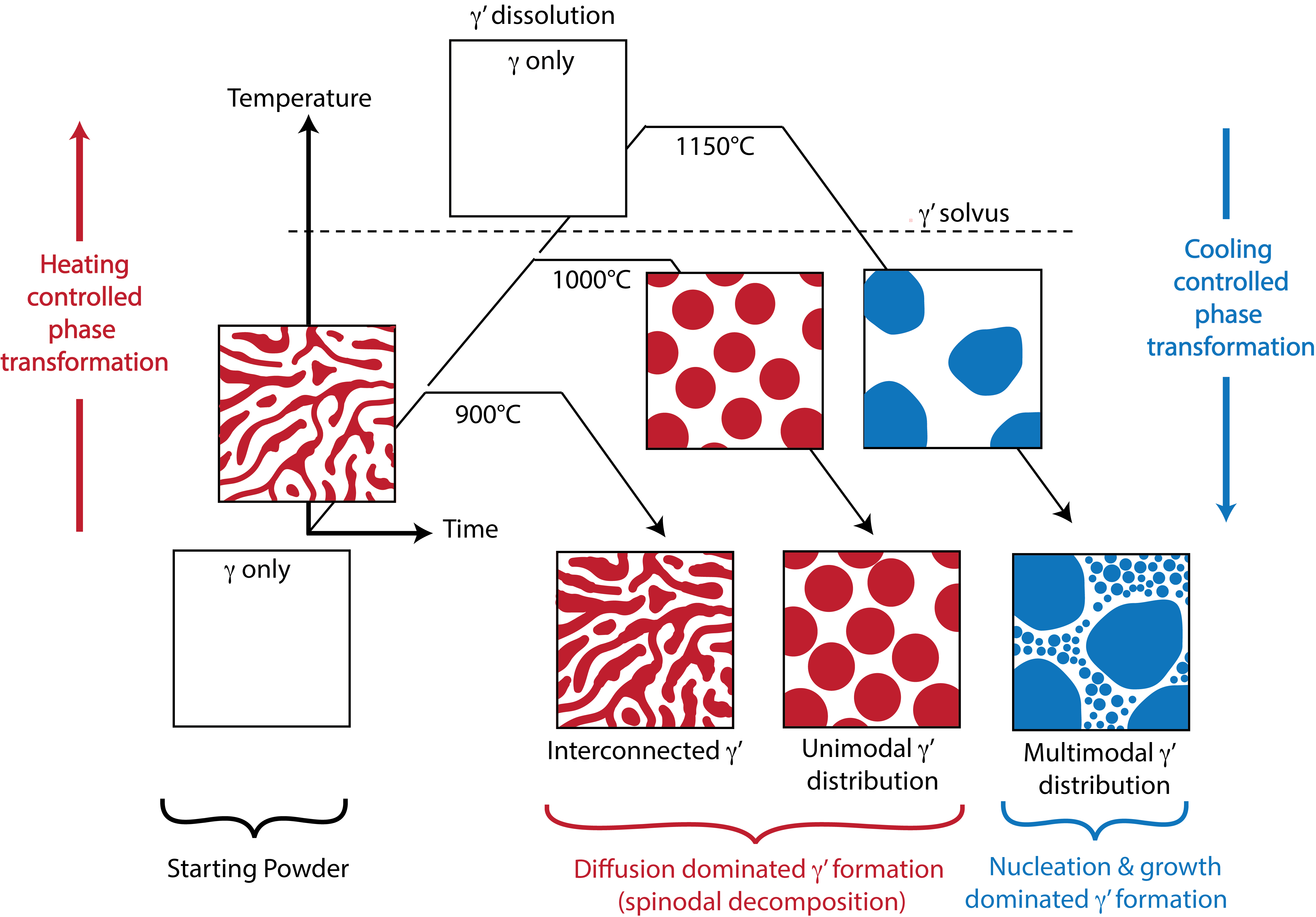}\\
	\caption{Schematic of $\gamma'$ formed via heating or cooling with their associated phase transformation mechanisms.}
	\label{Mechanism_schematic}
\end{figure*}

A schematic summary of the observed $\gamma'$ formation behaviour is given in Fig. \ref{Mechanism_schematic}. Notably, the microstructures following each of the heat treatments have significantly different characteristics. The samples that were exposed to the subsolvus temperatures are discussed first. The 900$^{\circ}$C sample leaves an interconnected network of $\gamma'$ precipitates, consistent with precipitates observed in superalloys \cite{VISWANATHAN2011485} and other systems \cite{CAHN1961795} where the formation was proposed to be via spinodal decomposition.

The starting $\gamma$ phase will undergo phase separation via spinodal decomposition, in which nm-scale domains within the $\gamma$ phase possess local chemical enrichment of certain elements such as Al, Ti, Ta and Ni, considered to be the $\gamma'$ forming elements, whilst they are lean in the elements such as Co, Cr and Mo. Forming domains which become enriched/depleted in these elements is easy as precursor clustering from the nm-scale segregation is already present in the powder following the gas atomisation. From the chemical fluctuations, an ordering reaction from the A1$\rightarrow$L$1_2$ structure will follow; noting that short range order is easy due to the small atomic displacements needed, leading to a fine scale dispersion of ordered $\gamma'$ precipitates. 

This reaction is thermodynamically favourable if, at a low temperature, a miscibility gap exists within the $\gamma$ phase where compositional partitioning through spinodal decomposition leads to the formation of two disordered A1 structured phases, $\gamma_{(1)}$ and $\gamma_{(2)}$. Subsequent chemical ordering within one of these disordered phases will result in  L1$_2$ structured $\gamma'$ precipitates. \cite{Tan2014, VISWANATHAN2011485}. The composition of this alloy is proposed to lie within this region where a spontaneous unmixing reaction is favourable. It is proposed that as the sample was heated, regions with chemical fluctuations were more stable than the $\gamma$ phase, leading to an energetically favourable phase separation to $\gamma_{(1)}$ + $\gamma_{(2)}$. The thermal energy by heating enables a continuous phase separation reaction to proceed, promoting the formation of Al/Ti/Ta/Ni rich domains to form. When the enrichment has reached a critical value, a favourable L$1_2$ chemical ordering reaction will ensue. This mechanism assumes that the transformation pathway follows spinodal decomposition-based phase separation first, followed by structural ordering; this sequence seems probable considering the segregation of Al/Ti/Ni and Co/Cr in the starting condition, and the L$1_2$ $\gamma'$ structure is detected via diffraction only when the compound exceeds $\sim$700$^\circ$C, this corresponds to the temperature where diffusion is favourable for significant solute transport. 

An alternative transformation pathway known as conditional spinodal \cite{Soffa1989} is also possible. In this mechanism, a congruent ordering reaction of the $\gamma$ phase occurs leading to the formation of a single ordered $\gamma'$ phase. Spinodal decomposition subsequently occurs within the $\gamma'$ phase to form two ordered L1$_2$ structured phases, $\gamma'_{(1)}$ and $\gamma'_{(2)}$. Only when either $\gamma'_{(1)}$ or $\gamma'_{(2)}$ undergoes a disordering reaction to form $\gamma$, will the final microstructure comprise both $\gamma$ + $\gamma'$. It is, however, noted that confirming this mechanism is operative during heating would be challenging as the equilibrium $\gamma$ phase fraction increases with temperature. Distinguishing this effect over a $\gamma'_{(1)} \rightarrow \gamma$ or $\gamma'_{(2)} \rightarrow \gamma$ disordering reaction, via neutron diffraction measurements as performed in this study, would be difficult.

Once the $\gamma'$ structure exists, the $\gamma'$ precipitates grow with a concomitant long-range diffusion and structural ordering, as evidenced by the in-situ measurements of increasing $\gamma'$ volume fraction. The $\gamma'$ morphology change and volume fraction increase is now driven by (1) minimising the elastic strain energy from lattice misfit strains (2) the reduction of the large interfacial energy associated with a fine scale $\gamma'$ and (3) reach the thermodynamic equilibrium composition.

\begin{figure}[ht!]
	\centering
	\includegraphics[width=80mm]{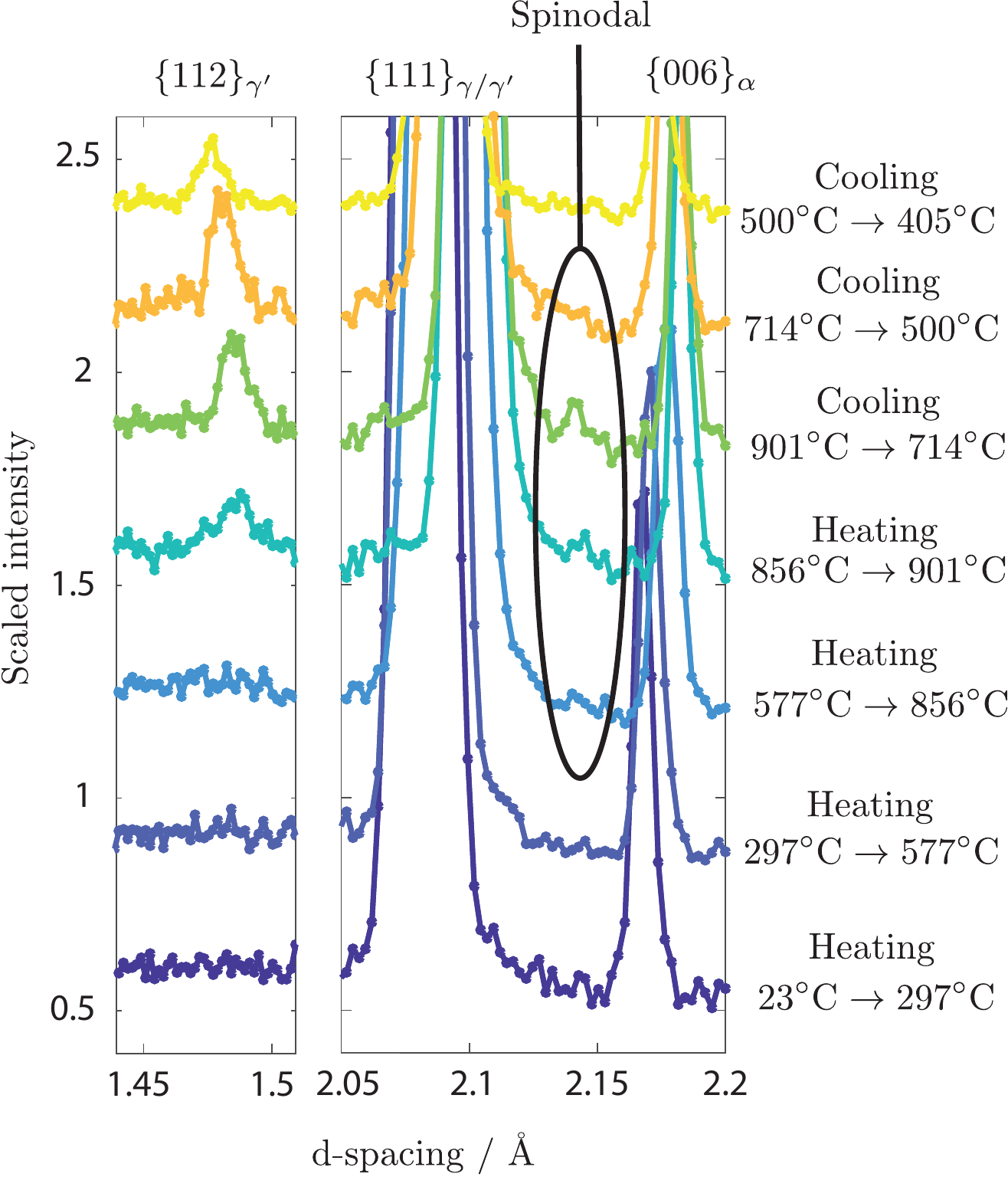}\\
	\caption{Identifying satellite reflections indicative of spinodal decomposition for the sample heated to 900$^\circ$C.}
	\label{SpinodalDiffractionFig}
\end{figure}

Further evidence of spinodal decomposition during heating is observed from the presence of small satellite reflections in the neutron diffraction data, close to fundamental reflections. The observation of satellite reflections are direct measurements of the composition modulation, as has been demonstrated on different systems via XRD \cite{PRASADRAO1987199} and TEM \cite{zhao1999}. Evidence of the satellite reflection is observed in the neutron diffraction data in this study, in the vicinity of the \{111\} reflection, as shown in Fig. \ref{SpinodalDiffractionFig}. These satellite reflections were very weak, however, could be observed when the data was summed. Each pattern in Fig. \ref{SpinodalDiffractionFig} corresponds to data acquired for 14 minutes. Whilst the temporal resolution here is not high, it is evident that the composition modulation associated with spinodal is evident during heating, isothermal hold and initial cooling. This satellite peak disappears as the sample cools, indicative that the composition fields are moving towards equilibrium, and the modulation is no longer strong enough to produce this diffraction characteristic.

As there is evidence that the sample heated to 1000$^{\circ}$C resulted in $\gamma'$ initially forming during heating, the resulting $\gamma'$ must have formed via spinodal decomposition, similar to the 900$^{\circ}$C sample. The resulting microstructure comprising an approximately unimodal distribution, not interconnected $\gamma'$ like the 900$^{\circ}$C sample, indicates this evolution is diffusion driven, enabled by the higher temperature heat treatment. During this time, the $\gamma'$ form morphologically stable spherical precipitates, as the interfacial energy of the precipitates reduces via coarsening, as expected from the Cahn-Hilliard model for spinodal decomposition \cite{CahnHilliard}. The unimodal $\gamma'$ size distribution further indicates all precipitates formed together; initiating upon heating. As the 1000$^{\circ}$C sample cools there is evidence of near zero lattice misfit, indicating the precipitates did not have to overcome energetically unfavourable misfit strains, which if present, would have promoted non-spherical precipitate morphologies due to the anisotropic stiffness of the cubic $\gamma'$ \cite{Ricks1983}.

\subsection{Precipitate formation during cooling}

When $\gamma'$ has formed during cooling, as is evident from the sample heated to a supersolvus temperature of 1150$^{\circ}$C then cooled, there is considerable evidence to indicate the precipitate formation mechanism is dissimilar to $\gamma'$ formed during heating. The multimodal distribution is consistent with observations made by others, who attribute the generation of $\gamma'$ from classical nucleation and growth \cite{WEN20031123, RADIS20095739, SINGH2011878}. Each generation of $\gamma'$ forms during cooling, where the driving force for $\gamma'$ nucleation occurs from increased undercooling. Growth of a $\gamma'$ distribution will continue whilst the temperature is high, though its driving force for solute diffusion decreases as temperature becomes lower. Growth is inhibited by soft impingement when solute fields overlap, however, further cooling results in non-equilibrium $\gamma'$ compositions away from the $\gamma/\gamma'$ interfaces, which with further undercooling, result in a further nucleation burst \cite{SINGH2013280}. In this study, this mechanism is corroborated by (1) the presence of secondary $\gamma'$ and tertiary $\gamma'$ precipitate distributions, (2) the compositions of the precipitate distributions are dissimilar, indicating their formation at different temperatures as the supersaturated $\gamma$ composition changes, (3) the dissimilar morphologies of secondary $\gamma'$ (octodendritic) and tertiary $\gamma'$ (spherical) indicates their growth has occurred with precipitates from each distribution subjected to different misfit strains. The initial secondary $\gamma'$ to form upon cooling has a lattice misfit of -0.5\%, sufficient to favour octodendritic morphology to develop due to the interaction between diffusion driven growth and non-uniform elastic strain fields around precipitates from the $\gamma'$ elastic anisotropy \cite{Ricks1983}. This effect diminishes as the temperature is lowered, such that the lattice misfit is close to zero when the tertiary $\gamma'$ forms, resulting in spherical precipitates for this distribution. Due to these factors, there was no evidence that the $\gamma'$ precipitates formed by spinodal decomposition during cooling.

\subsection{Implications}

The $\gamma'$ that has formed during heating, in the 900$^{\circ}$C and 1000$^{\circ}$C samples, is entirely a diffusion driven process. A classical nucleation is not plausible as there is no undercooling, and therefore no driving force for this process. Unimodal $\gamma'$ size distributions, similar to the microstructure achieved here from the sample heated to 1000$^{\circ}$C, have been achieved by others by cooling samples rapidly \cite{BABU20014149, Hwang2009, COLLINS201496}. There remains disagreement in the literature on how such distributions form upon cooling; where the precipitates are nucleation controlled, though undercooling is so high that it outruns diffusion and the condition for soft impingement is not met \cite{WEN20031123}, whereas Viswanathan \textit{et al.} \cite{VISWANATHAN2011485} indicates phase separation initiates by spinodal decomposition. There is clear scope to widen the study, particularly for different compositions and cooling rates to confirm under the precise conditions spinodal decomposition mechanism prevails. 

Whilst unimodal microstructures created by rapid cooling offer desirable increases in yield strength over slower cooled samples \cite{COLLINS201496}, quenching is impractical for real manufacture of large components. Problems include the creation of high residual stresses \cite{Bi2014}, and transient thermal gradients and inhomogeneous microstructures \cite{ZHANG2019107603}. The study here creates similarly desirable microstructures (in the 1000$^{\circ}$C sample, for example) directly from the powder without cooling related issues. Control at this stage could be used to mitigate or change parameters of subsequent thermomechanical treatments. This work also has enormous implications for the use of nickel-base superalloy powders used in additive manufacturing (AM) processes, where the $\gamma \rightarrow \gamma'$ transformation behaviour is largely unknown. Such AM process could exploit microstructural control through $\gamma'$ formation during heating regimes for the first time.

Finally, in context to the HIP treatment of powders during the commercial fabrication of turbine rotor discs, this study gives insight into the microstructural development. The trimodal $\gamma'$ distribution comprising primary, secondary and tertiary $\gamma'$ during subsolvus HIP treatments of RR1000  \cite{QIU2013176} was not observed here. The absence of pressure in this study prevents primary $\gamma'$ forming. This is to be expected as elevated temperature alone enables $\gamma'$ to form before powder particles are fully consolidated. It is speculated that only when an externally applied pressure is sufficient to fully consolidate powder particles before $\gamma'$ formation begins, diffusional processes at grain boundaries required to nucleate primary $\gamma'$ are possible. Confirming this mechanism is proposed for future work.

\subsection{Characterisation}

To quantify the $\gamma \rightarrow \gamma'$ reactions in-situ, the ENGIN-X beamline was selected for its ability to obtain diffraction data from a large sample volume (several mm$^3$), without the signal being significantly attenuated by an alumina crucible containing the powder, a high temperature furnace and a rapid data acquisition rate (1.5 minutes per pattern). To analyse results from data collected at this rate, however, a new data analysis strategy was employed here, based on constrained Gaussian profiles. 

The diffraction line profile shape used here was Gaussian, deemed to be a suitable approximation to the experimentally obtained line profile shape. Whilst the pure instrument profile for reflections from ENGIN-X is better fitted with a convolution of a Voigt function with a truncated exponential \cite{Santisteban:ks5105}, no peak asymmetry nor Cauchy character was  evident, which is explained by dominant broadening contributions from the specimen geometry and sample size/strain broadening. Using a Gaussian line profile also had the advantage that the intensities of the $\gamma$ and $\gamma'$ reflections of a fundamental reflection can be easily coupled via the analytical solution of integrated Gaussian intensity (i.e. Equation \ref{GaussCouple}). This significantly constrained the problem of fitting 2 superimposed reflections with independent variables for each, which is most difficult when $\gamma$ and $\gamma'$ phases have near identical lattice parameters. Fitting the data here with more complex functions, such as those with convolutions, was found to be inappropriate due to increased fitting variables that results in scatter in the fitted coefficients, primarily due to numerical instability. The approach adopted here demonstrated that the use of more complex fitting functions to analyse superalloy datasets is suitable only when high signal-to-noise data is available (i.e. synchrotron data over time-of-flight data collected here) or when $\gamma/\gamma'$ differences in lattice parameter make direct peak separation possible.

\section{Conclusions}

The phase transformation from $\gamma$ to $\gamma'$ was investigated in a nickel-base superalloy powder subjected to heat treatments at 900$^{\circ}$C, 1000$^{\circ}$C and 1150$^{\circ}$C for 15 min with controlled heating and cooling during in-situ neutron diffraction. Insights here were combined with 
post mortem characterisation using atom probe tomography and electron microscopy to elucidate composition and microstructure evolution. From this, the following conclusions can be drawn: 
\begin{enumerate}
		\item{Characterisation of the gas atomised powder, prior to any further processing, shows no evidence of structural ordering, however, near-atomic scale compositional segregation was observed. Cr and Co atoms were found to segregate together, whereas Ni, Al and Ti atoms are repelled by Cr atoms resulting in regions enriched in $\gamma'$ elements. This segregational behaviour can act as the precursor to phase separation.}
		\item{A thermal cycle reaching 900$^{\circ}$C leads to the formation of $\gamma'$ precipitates during heating by spinodal decomposition, confirmed by neutron diffraction and atom probe tomography. The $\gamma'$ precipitates in this case formed only during heating and possessed an interconnected $\gamma'$ network with an irregular morphology. No $\gamma'$ precipitates formed during cooling. Significant composition variation was observed in the $\gamma$ phase in the post mortem analysis.}
		\item{A thermal cycle reaching 1000$^{\circ}$C similarly formed $\gamma'$ via spinodal decomposition during heating. The $\gamma'$ volume fraction increased during cooling whilst solute diffusion prevailed, as the system approached thermodynamic equilibrium. Observing this material post mortem indicates the system is indeed closer to equilibrium compared to the 900$^{\circ}$C sample, comprising a unimodal $\gamma'$ distribution, a homogenous $\gamma$ \& $\gamma'$ composition and coarser spherical precipitate morphology, indicative of a reduced interfacial energy.}
		\item{A multimodal $\gamma'$ distribution was observed after heating above the $\gamma'$ solvus at 1150$^{\circ}$C, where $\gamma'$ precipitates formed only during cooling based on classical nucleation and growth mechanisms; there was no evidence of spinodal decomposition. This mechanism was corroborated by distribution of precipitates sizes, compositional differences between different populations of secondary and tertiary $\gamma'$ precipitates, and morphology differences between the $\gamma'$ distributions associated with their different lattice misfit strains during their formation. Chemical fluctuations were also observed in the $\gamma$ matrix as a function of distance from secondary $\gamma'$ precipitates.}

\end{enumerate}

\section*{Acknowledgements}
The authors would like to gratefully acknowledge Rolls-Royce for the provision of material and the Science and Technology Facilities Council (STFC) for access to neutron beamtime at ISIS, the experimental assistance from Dr Joe Kelleher, and also for the provision of sample preparation, ENGIN-X facilities. D.C. acknowledges financial support from his Birmingham Fellowship. C.P. would like to acknowledge the funding from Innovation Fellowship (EP/S000828/1) funded by Engineering and Physical Science Research Council (EPSRC), UK Research and Innovation. P.K. acknowledges financial support from the DFG SFB TR 103 through project A4.
\section*{References}

\bibliographystyle{model3-num-names}

\newpage

\end{document}